# Stimulated emission and absorption of radiation in a single mode (N-TLM mode) for various photon distributions

Michael Tavis

## Abstract

The Tavis-Cummings model of N two-level atoms interacting with a single resonant mode is extended to various cases of off-resonance, initial photon densities, and atom number N. For stimulated absorption and small numbers of mean photon number a very unexpected results showing well defined rise and fall of oscillation is found.

## Background:

The time development of the ensemble averages of $E^{-}$ and $E^{-}E^{+}$ for photon number densities representing the coherent and thermal states [1] was given in 1965 but only for the resonant case and short times [2]. Eberly [3] provided the long-term development for the ensemble average of $E^{-}E^{+}$ showing the periodic collapse and revival of oscillations in that average. An elegant empirical demonstration of the quantum collapse and revival was observed in 1987, when a single Rydberg atom in a single mode of an electromagnetic field in a superconducting cavity was investigated [4]. Since that time, numerous papers have been published on the subject of quantum radiation in interaction with atoms. More recently, results for simulated emission in a single mode were given for both resonance and non-resonance for various initial photon density distributions [5]. In1968, the exact solutions to what has become known as the Tavis-Cummings Model were presented [6,7], extending to N atoms the basic model of one atom originally proposed in 1963 to study the relationship between the quantum theory of radiation and the semi-classical theory in describing spontaneous emission [8].

In the following, we will apply the solutions of the Tavis-Cummings model for N two-level molecules (N-TLMs) as basis states to find expressions for the time development of the ensemble averages of $E^{-}$ and $E^{-}E^{+}$. With the general solution in hand, specific cases of interest will be presented in more detail for various photon distributions previously developed [5].

A revision to the original submission is made for this submission. Eberly[3] defined a photon revival time defined as $\tau_{R} = 2\pi\frac{\sqrt{\bar{n}+1+\Delta}}{\gamma}$ where Δ is a measure of non-resonance and will be defined later. We will use this revival time to renormalize time in the following numerical examples. The reason for using normalized time is that it reveals behavior not obvious for non-normalized time. We will discuss those observation later.

### Ensemble Averages for $\boldsymbol{E^{-}}$ and $\boldsymbol{E^{-}E^{+}}$

The ensemble average of the field operators $E^{-}(t)$ and $E^{-}E^{+}(t)$ may be found in the usual way [9]

$$\langle E^{-}(t)\rangle = -\left(\frac{\gamma}{\mu}\right)\sum_{n=0}^{\infty}(n+1)^{1/2}\langle n|\rho_{f}(t)|n+1\rangle, \tag{1a}$$

and

$$\langle E^{-}E^{+}(t)\rangle = \left|\frac{\gamma}{\mu}\right|^{2}\sum_{n=0}^{\infty} n\langle n|\rho_{f}(t)|n\rangle, \tag{1b}$$

where only one mode of the field is excited, $\gamma$ is the complex coupling constant, and $\mu$ is the dipole moment of the TLM with which the field is interacting. The elements of the field density matrix are given by the trace over the TLM states

$$\langle n|\rho_{f}(t)|n'\rangle = \sum_{r,m} P(r)\,\langle n|\langle r,m|\rho(t)|r,m\rangle|n'\rangle \tag{2}$$

where

$$P(r) = \frac{N!\,(2r+1)}{\left(\frac{N}{2}+r+1\right)!\left(\frac{N}{2}-r\right)!} \tag{3}$$

and $\rho(t)$ is given by a unitary transformation of the density operator at time $t_0 = 0$ where it is assumed that the N-TLMs and radiation field are not interacting! Therefore

$$\rho(t) = U(t)\rho(0)U^{-1}(t), \tag{4}$$

$$U(t) = e^{iHt}, \tag{5}$$

and where H is given [6,7] [1] by

$$H = (a^{\dagger}a + R_3) + a^{\dagger}a\,\frac{\omega-\Omega}{\Omega} - \kappa a R_{+} - \kappa^{*}a^{\dagger}R_{-} \tag{6}$$

Since the system is non-interacting at time zero, the density operator is a direct product of the field part and N-TLM part of the system.

$$\rho(0) = \rho_{TLM} \otimes \rho_{f.} \tag{7}$$

Equation (2) can be expanded directly in terms of the orthonormal eigenvectors of the Hamiltonian above assuming that all the TLMs are at equivalent mode positions, namely the |r,c,j> states which are expressed in [6],

$$|r,c,j> = \sum_{n=Max[0,c-r]}^{c+r} A_{n}^{r,c,j}|n>|r,c-n>. \tag{8}$$

---

[1] Reference 7 contains the full development for both resonance and non-resonance while reference 6 contains only the resonance results for which the photon energy $\omega$ is equal to the energy separation $\Omega$ for the TLM.

In expression (8), r and c are good (conserved) quantum numbers of the Hamiltonian where $r \leq \frac{N}{2}$ and c=n+m with $-r \leq m \leq r$. Then

$$\begin{aligned}\langle n|\rho_f(t)|n'\rangle = \sum_{r,m} P(r) \sum_{r',c',j'} \sum_{r'',c'',j''} \langle n|\langle r,m|e^{-iHt}|r',c',j'\rangle \\ \times \langle r',c',j'|\rho(0)|r'',c'',j''\rangle\langle r'',c'',j''|e^{iHt}|r,m\rangle|n'\rangle.\end{aligned} \tag{9}$$

With this and a fair amount of effort [10], equations (1a) and (1b) can be written as[2]

$$\begin{aligned}
\langle E^-(t)\rangle &= -\left(\frac{\gamma}{\mu}\right)\sum_{n=0}^{\infty}(n+1)^{1/2}\langle n|\rho_f(t)|n+1\rangle \\
&= -\left(\frac{\gamma}{\mu}\right)\sum_{r=0,\frac{1}{2}}^{\frac{N}{2}} P(r)\sum_{m=-r}^{r}\sum_{c=m}^{\infty}(c-m+1)^{1/2}\sum_{p=-r}^{r}\sum_{j=0}^{Min[2r,c+r]}\sum_{p'=-r}^{r}\sum_{j'=0}^{Min[2r,c+r+1]}(A^*)_{c-p}^{r,c,j} \\
&\quad \times A_{c-m}^{r,c,j}(A^*)_{c-m+1}^{r,c+1,j'}A_{c-p'+1}^{r,c+1,j'}e^{-i\lambda_{r,c,j}t}e^{i\lambda_{r,c+1,j'}t}\langle c-p|\rho_f(0)|c+1-p'\rangle\,\langle r,p|\rho_{TLM}(0)|r,p'\rangle, \\
\langle E^-E^+(t)\rangle &= \left|\frac{\gamma}{\mu}\right|^2\sum_{n=0}^{\infty}n\langle n|\rho_f(t)|n\rangle \\
&= \left|\frac{\gamma}{\mu}\right|^2\sum_{r=0,\frac{1}{2}}^{\frac{N}{2}} P(r)\sum_{m=-r}^{r}\sum_{c=m}^{\infty}(c-m)\sum_{p=-r}^{r}\sum_{j=0}^{Min[2r,c+r]}\sum_{p'=-r}^{r}\sum_{j'=0}^{Min[2r,c+r]}(A^*)_{c-p}^{r,c,j} \\
&\quad \times A_{c-m}^{r,c,j}(A^*)_{c-m}^{r,c,j'}A_{c-p'}^{r,c,j'}e^{-i\lambda_{r,c,j}t}e^{i\lambda_{r,c,j'}t}\langle c-p|\rho_f(0)|c-p'\rangle\,\langle r,p|\rho_{TLM}(0)|r,p'\rangle,
\end{aligned} \tag{10}$$

As is seen, these equations are quite complicated and to use the general expressions would be prohibitive except for the simplest cases. This completes the formulation of the general expressions. We only consider two special cases next.

## Special cases

Equation (10) can be simplified considerably by considering special cases for the initial TLM distribution. Towards that end, we consider cases for simulated emission and absorption where all the TLMs are initially in the up state or down state.

### All TLMs initially in the up state-Simulated Emission

For this case $r = \frac{N}{2}$ and $P(r) = 1$. Note that r is the cooperation number, n the photon number and c=n+r. Further

$$\langle r,p|\rho_{TLM}(0)|r,p'\rangle = \delta_{p,\frac{N}{2}}\delta_{p',\frac{N}{2}} \tag{11}$$

Equation (10) becomes

---

[2] The factors $\left(\frac{\gamma}{\mu}\right)$ and $\left|\frac{\gamma}{\mu}\right|^2$ seen in eq. (10) differ from those same constants in reference [2] by a factor of 2, i.e. $\gamma$ *is* $2\gamma$ in that reference. We ignore the difference here.

$$
\begin{aligned}
\langle E^{-}(t)\rangle &= -\left(\frac{\gamma}{\mu}\right) \sum_{m=-\frac{N}{2}}^{\frac{N}{2}} \sum_{c=m}^{\infty} (c-m+1)^{1/2} \langle c-\frac{N}{2}\left|\rho_f(0)\right| c+1-\frac{N}{2}\rangle \\
&\quad \times \sum_{j=0}^{Min\left[N,c+\frac{N}{2}\right]} \sum_{j'=0}^{Min\left[N,c+1+\frac{N}{2}\right]} e^{-i\left(\lambda_{\frac{N}{2},c,j}-\lambda_{\frac{N}{2},c+1,j'}\right)t} (A^{*})_{c+1-\frac{N}{2}}^{\frac{N}{2},c+1,j} A_{c-m}^{\frac{N}{2},c,j} (A^{*})_{c-m}^{\frac{N}{2},c,j'} A_{c+1-\frac{N}{2}}^{\frac{N}{2},c+1,j'} \\
\langle E^{-}E^{+}(t)\rangle &= \left|\frac{\gamma}{\mu}\right|^{2} \sum_{m=-\frac{N}{2}}^{\frac{N}{2}} \sum_{c=m}^{\infty} (c-m) \langle c-\frac{N}{2}\left|\rho_f(0)\right| c-\frac{N}{2}\rangle \\
&\quad \times \sum_{j=0}^{Min\left[N,c+\frac{N}{2}\right]} \sum_{j'=0}^{Min\left[N,c+\frac{N}{2}\right]} e^{-i\left(\lambda_{\frac{N}{2},c,j}-\lambda_{\frac{N}{2},c,j'}\right)t} (A^{*})_{c-\frac{N}{2}}^{\frac{N}{2},c,j} A_{c-m}^{\frac{N}{2},c,j} (A^{*})_{c-m}^{\frac{N}{2},c,j'} A_{c-\frac{N}{2}}^{\frac{N}{2},c,j'}
\end{aligned}
\tag{12}
$$

In addition, the density matrices $\langle c-\frac{N}{2}\left|\rho_f(0)\right|c-\frac{N}{2}\rangle$ and $\langle c-\frac{N}{2}\left|\rho_f(0)\right|c+1-\frac{N}{2}\rangle$ only have values for $c-\frac{N}{2}\geq 0$. By changing variables and using the orthogonality relationships for $A_n^{r,c,j}$ the following expressions can be found [10]:

$$
\langle E^{-}E^{+}(t)\rangle = \left|\frac{\gamma}{\mu}\right|^{2} [\bar{n} + S_1(\bar{n},N,\gamma t)],
\tag{13}
$$

where $\bar{n}$ is the mean number of photons and

$$
\begin{aligned}
S_1(\bar{n},N,\gamma t) &= -4 \sum_{n=0}^{\infty} \langle n\left|\rho_f(0)\right|n\rangle \sum_{j=0}^{N-1} \sum_{j'=j+1}^{N} Sin^2\left\{\frac{\left[q_{\frac{N}{2},\left(n+\frac{N}{2}\right),j} - q_{\frac{N}{2},\left(n+\frac{N}{2}\right),j'}\right]}{2} \Omega|\kappa| t\right\} \\
&\quad \times (A^{*})_{n}^{\frac{N}{2},\left(n+\frac{N}{2}\right),j} A_{n}^{\frac{N}{2},\left(n+\frac{N}{2}\right),j'} \sum_{p=0}^{N} p\, A_{n+p}^{\frac{N}{2},\left(n+\frac{N}{2}\right),j} (A^{*})_{n+p}^{\frac{N}{2},\left(n+\frac{N}{2}\right),j'}
\end{aligned}
\tag{14}
$$

Equation (14) is the final general solution for $S_1(\bar{n},N,\gamma t)$ (no assumptions about resonance or non-resonance) and we have used the alternate expression for the effective eigenvalues of reference [6], namely λ=c-$|\kappa|q$ and multiplied by $\Omega$ to obtain the correct units even though ħ is still set to unity. Note that $\Omega|\kappa|t = \gamma t$. Using the notation introduced by reference [2], the ensemble average for the field $\langle E^{-}(t)\rangle$ can be written as

$$
\langle E^{-}(t)\rangle = -\left(\frac{\gamma}{\mu}\right) e^{i\omega t} S_2(\bar{n},N,\gamma t)
\tag{15}
$$

where

$$S_2(\bar{n}, N, \gamma t) = \sum_{n=0}^{\infty} \langle n|\rho_f(0)|n+1\rangle \sum_{k=0}^{N} (n+k+1)^{1/2}$$
$$\times \left\{ \sum_{j=0}^{N}\sum_{j'=0}^{N} Cos\left[\left(q_{\frac{N}{2},n+\frac{N}{2},j} - q_{\frac{N}{2},n+\frac{N}{2}+1,j'} - \beta\right)\Omega|\kappa|t\right] (A^*)_{n+1}^{\frac{N}{2},n+\frac{N}{2}+1,j} A_{n+k}^{\frac{N}{2},n+\frac{N}{2},j} (A^*)_{n+k}^{\frac{N}{2},n+\frac{N}{2},j'} A_{n+1}^{\frac{N}{2},n+\frac{N}{2}+1,j'} \right.$$
$$\left. + i \sum_{j=0}^{N}\sum_{j'=0}^{N} Sin\left[\left(q_{\frac{N}{2},n+\frac{N}{2},j} - q_{\frac{N}{2},n+\frac{N}{2}+1,j'} - \beta\right)\Omega|\kappa|t\right] (A^*)_{n+1}^{\frac{N}{2},n+\frac{N}{2}+1,j} A_{n+k}^{\frac{N}{2},n+\frac{N}{2},j} (A^*)_{n+k}^{\frac{N}{2},n+\frac{N}{2},j'} A_{n+1}^{\frac{N}{2},n+\frac{N}{2}+1,j'} \right\} \quad (16)$$

and we have taken advantage of the real nature of the $A_n^j$. The relative tuning parameter is a measure of non-resonance and is given by $\beta = \frac{\omega - \Omega}{|\kappa|\Omega}$. Simplification of $S_2$ is difficult except for resonance. We will not consider $\langle E^-(t)\rangle$ or $S_2(\bar{n}, N, \gamma t)$ further in this paper. Specific examples for N=1 through 4 for resonance and N=1 and 2 for non-resonance were provided in reference [10].

***Numerical evaluation of*** $\boldsymbol{S_1(\bar{n}, N, \tau)}$

The exact calculation for this expression is performed using eq. (14). In order to perform these calculations, we make use of the photon number densities developed in reference [5], namely Eqs. 17-28 in Table 1.

| Table 1: Photon Number Distributions | | |
|---|---|---|
| Coherent State[3] | $\rho_{nn} = \frac{e^{-\beta^2}\|\beta\|^{2n}}{n!}, \bar{n} = \|\beta\|^2, var = \|\beta\|^2$ | 17 |
| Thermal State[6] | $\rho_{nn} = \frac{1}{\bar{n}_T + 1}\left(\frac{\bar{n}_T}{\bar{n}_T + 1}\right)^n, \bar{n} = \bar{n}_T, var = {\bar{n}_T}^2 + \bar{n}_T$ | 18 |
| Fock State | $\rho_{nm} = \delta_{nm}, \quad \bar{n} = n, var = 0$ | 19 |
| Mixed Coherent and Thermal State[2] | $\rho_{nm} = \frac{e^{-\frac{\beta^2}{\bar{n}_T+1}}}{(\bar{n}_T + 1)}\left(\frac{\bar{n}_T}{\bar{n}_T + 1}\right)^n M\left(-n, 1, \frac{-\beta^2}{\bar{n}_T(\bar{n}_T + 1)}\right),$ <br> $\bar{n} = \|\beta\|^2 + \bar{n}_T, \quad var = \|\beta\|^2(1 + 2\bar{n}_T) + {\bar{n}_T}^2 + \bar{n}_T$ | 20 |
| Squeezed Vacuum State[4] | $\rho_{nm} = \frac{\left[\frac{1}{2}Tanhr\right]^n n!}{Coshr\left(\frac{n}{2}!\right)^2} \quad n\ even, \quad \bar{n} = Sinh^2 r$ <br> $= 0 \quad n\ odd, \quad var = \frac{Sinh^2(2r)}{2}$ | 21 |
| Squeezed Fock State[7] for state $l$ | $\rho_{nn}(l) = \frac{l!\,n!}{(Coshr)^{2n+1}}\left(\frac{1}{2}Tanhr\right)^{l-n} S(r,n,l)\ \|n - l\|\ even, \quad \bar{n} = l + (2l+1)Sinh^2 r$ <br> $= 0 \quad \|n - l\|$ odd, $\quad var = \frac{1}{2}(l^2 + l + 1)Sinh^2(2r)$ | 22 |
| Squeezed Thermal State[7] | $\rho_{nn} = \frac{1}{\bar{n}_T + 1}\sum_{l=0}^{\infty} \rho_{nn}(l)\left(\frac{\bar{n}_T}{\bar{n}_T + 1}\right)^l,$ <br> $\bar{n} = \bar{n}_T + (2\bar{n}_T + 1)Sinh^2 r, var = -\frac{1}{4} + \left(\bar{n}_T + \frac{1}{2}\right)^2 Cosh(4r)$ | 23 |

---

[3] Roy J. Glauber, Coherent and Incoherent States of the Radiation Field, Physical Review, Vol. 131, #6, Sept. 1963, p. 2766.

[4] M. S. Kim, F. A. M. de Oliveira, and P. L. Knight; Properties of squeezed number states and squeezed thermal states; Physical Review A, Vol. 40, # 5, Sept. 1, 1989

| Table 1: Photon Number Distributions | | |
|---|---|---|
| Squeezed Coherent State[5] | $\rho_{nn} = N_1(\beta,r,\psi)\frac{\left[\frac{1}{2}Tanh(r)\right]^n}{n!}\left\lvert H_n\left\{\frac{\lvert\beta\rvert}{\sqrt{2}}\left[e^{-i\frac{\psi+\pi}{2}}Tanh^{-\frac{1}{2}}(r)+e^{i\frac{\psi+\pi}{2}}Tanh^{\frac{1}{2}}(r)\right]\right\}\right\rvert^2$<br>$\bar{n} = Sinh^2r + \lvert\beta\rvert^2, var = \lvert\beta\rvert^2[Cosh(2r)+Cos(\psi)Sinh(2r)] + \frac{Sinh^2(2r)}{2}$ | 24 |
| Mixed Squeezed Coherent State and Thermal State[6] | $\rho_{nn} = N_2(\beta,\bar{n}_T,r)\frac{1}{n!}\left[\frac{\bar{n}_T}{1+\bar{n}_T}\right]^n H_{n,n}(r_1,r_2)$<br>$\bar{n} = Sinh^2r + \lvert\beta\rvert^2 + \bar{n}_T$<br>$var = \lvert\beta\rvert^2[Cosh(2r)+Cos(\psi)Sinh(2r)] + \frac{Sinh^2(2r)}{2} + 2\bar{n}_T(Sinh^2r+\lvert\beta\rvert^2) + {\bar{n}_T}^2 + \bar{n}_T$ | 25 |
| Displaced Squeezed Thermal State (DSTS)[7] | $\rho_{nn} = \pi Q(0)\tilde{A}^n\sum_{q=0}^{n}\frac{1}{q!}\binom{n}{q}\left(\frac{\lfloor\tilde{B}\rfloor}{2\tilde{A}}\right)^q\left\lvert H_q\left(\frac{\tilde{C}}{\sqrt{2\tilde{B}}}\right)\right\rvert^2$<br>$\bar{n} = \bar{n}_T + (2\bar{n}_T+1)Sinh^2r + \lvert\beta\rvert^2,$<br>$var = -\frac{1}{4} + \lvert\beta\rvert^2(1+2\bar{n}_T)[Cosh(2r)+Cos(\psi)Sinh(2r)] + \left(\bar{n}_T+\frac{1}{2}\right)^2 Cosh(4r)$ | 26 |
| Displaced Number (Fock) State[8] for state $l$ | $\rho_n(l) = \begin{cases} \frac{n!}{l!}\lvert\beta\rvert^{2(l-n)}e^{-\lvert\beta\rvert^2}\lvert\mathcal{L}_n^{l-n}\rvert^2 = \frac{\lvert\beta\rvert^{2(l-n)}e^{-\lvert\beta\rvert^2}}{n!\,l!}\left\lvert\sum_{k=0}^{n}\frac{n!\,l!\,(-1)^k\lvert\beta\rvert^{2(n-k)}}{k!\,(n-k)!\,(l-k)!}\right\rvert^2 & l \geq n \\ \frac{l!}{n!}\lvert\beta\rvert^{2(n-l)}e^{-\lvert\beta\rvert^2}\left\lvert\mathcal{L}_l^{n-l}\right\rvert^2 = \frac{\lvert\beta\rvert^{2(n-l)}e^{-\lvert\beta\rvert^2}}{n!\,l!}\left\lvert\sum_{k=0}^{l}\frac{n!\,l!\,(-1)^k\lvert\beta\rvert^{2(l-k)}}{k!\,(n-k)!\,(l-k)!}\right\rvert^2 & n > l \end{cases}$<br>$\bar{n} = l + \lvert\beta\rvert^2, \qquad var = \overline{n^2} - \bar{n}^2 = (2l+1)\lvert\beta\rvert^2$ | 27 |
| Squeezed Displaced Number State[9] for initial state m | $\rho_{nn} = \left\lvert\langle n\rvert\beta,m\rangle_g\right\rvert^2 = \frac{n!}{m!\,Cosh(r)}\left[\frac{Tanh(r)}{2}\right]^{m+n}$<br>$\times\ Exp\{-\lvert\beta\rvert^2[1-Cos(\psi)Tanh(r)]\}\left\lvert\sum_{i=0}^{min(m,n)}\frac{\binom{m}{i}}{(n-i)!}\left[-\frac{4}{Sinh^2(r)}\right]^{\frac{i}{2}}SS(i,m,n,\lvert\beta\rvert^2,r,\psi)\right\rvert^2$<br>$\bar{n} = \langle n\rangle = \lvert\beta\rvert^2 + (2m+1)Sinh^2(r) + m$<br>$var = \langle(\varDelta n)^2\rangle = \lvert\beta\rvert^2[Cosh(2r)+Cos(\psi)Sinh(2r)](2m+1) + \frac{1}{2}(m^2+m+1)Sinh^2(2r)$ | 28 |

See reference [5] for definitions of various functions used in Eqs. 17-28. In the following examples, we will not focus on numerical solutions for N=1 TLM since those cases were examined in some detail for the various photon densities in reference [5]. We will provide some numerical solutions for the photon densities represented by Eqs. 17-28 for N=10 as a reminder of what we considered in references [10, 15]. The photon density representative by Eq. 19 will not be considered. All of the examples provided in the Table 2 below are for 10 TLMs since calculation for larger values takes considerable time. Each row of the table contains the photon density on the left and the value of $\boldsymbol{S_1(\bar{n}, N, \tau)}$ as a function of time $\boldsymbol{\tau}$ on the right**.** Again $\bar{n}$ is the mean photon number and N the number of TLMs. In some cases, we provide only resonant cases. For others we provide only non-resonant cases such as for the thermal state and squeezed thermal state. It is noted that numerous examples could be provided for various values of coherent photon number, thermal photon number, squeezing parameter, phase, and initial photon number and for various off resonant parameters. The range of possible examples is too vast to explore in this paper, thus I provide the Mathematica notebook used to perform the calculations. Note 2: In order to make use of the normalized time as defined by Eberly, the Sin function in Eq. 14 is rewritten as $Sin^2\left\{\pi\left[q_{\frac{N}{2},\left(n+\frac{N}{2}\right),j} - q_{\frac{N}{2},\left(n+\frac{N}{2}\right),j'}\right]\sqrt{\bar{\mathrm{n}}+1+\Delta\tau}\right\}$, where $\bar{\mathrm{n}}$ is the mean number of photons and $\tau$ the normalized time.

---

[5] J. J. Gong and P. K. Aravind, Expansion coefficients of a squeezed coherent state in the number state basis, The American Journal of Physics, Vol. 58, Issue 10 Oct. 1990

[6] A. Vourdas, Superposition of Squeezed Coherent States with Thermal Light, Phy. Rev A Vol. 34, #4, Oct. 1986, p. 2366. Note that the expressions for the mean and variance were not given in this reference. In fact, the authors have not found this expression for this case in any reference.

[7] Paulina Marian and Tudor A. Marian, Squeezed States with Thermal Noise. I Photon-Number Statistics, Phy. Rev. A, Vol. 47, #5, May 1993, p. 4474.

[8] F. A. M. de Oliveira, M. S. Kim, P. L. Knight and V. Bužek, Properties of displaced number states, Phy. Rev. A, Vol. 41, #5, p2645

[9] P. Král, Displaced and Squeezed Fock states, Journal of Modern Optics, Vol. 37, #5, p889, 1990.

I did explore a small range of coherent photon number for the coherent state represented by Eq. 17 since this was an easily calculated example. When the number of photons is equal to or smaller than the number of TLMs, we see from the examples below that the photon number is insufficient to drive the TLMs to a state of equal up and down TLMs and that the reoccurring oscillations rapidly approach random oscillation.

It is believed that the rise and fall of oscillations seen for the non-resonant cases is due to the distortion of bases states away from the more or less harmonic state forms exhibited for the resonant cases. This distortion can be extreme to the extent of favoring only a few of the $A_n^{r,c,j}$ at the lower or upper values of n.

In Table 3 below, we provide examples which examine the effect of increasing photon number for the resonant coherent case represented by Eq. 17. Two sets are shown, one for 1 TLM and the other for 10 TLMs. As in Table 2, Table 3 is in a row format with the photon density on the left and the value of $\boldsymbol{S_1}(\boldsymbol{\bar{n}}, \boldsymbol{N}, \tau)$ as a function of time $\tau$ on the right**.** The rows alternate with one row for 1 TLM and the second for 10 TLMs. We see that as the photon number increases greatly beyond the number of TLMs, that the rise and fall becomes considerably sharper and further apart. This is attributed to the linearity of the differences of eigenvalues in Eq. 14.

By examining the exact solution found in Ref. [5] displayed in Eq. 29 for one TLM, we see that as $\bar{n}$ increases we can approximate Eq. 29 by the following equation 30 where $\bar{n}$ is the standard deviation in photon number.

$$S_1(\bar{n}, \boldsymbol{\tau}) = \sum_{n=0}^{\infty}\left(\frac{e^{-\bar{n}}\bar{n}^n}{n!}\right) sin^2\left[2\,\pi\left((\bar{n}+1)(n+1)\right)^{1/2}\tau\right] \qquad 29$$

$$S_1(\bar{n}, \boldsymbol{\tau}) \simeq \sum\nolimits_{0}^{2\bar{n}}\left(\frac{e^{-\bar{n}}\bar{n}^n}{n!}\right) sin^2\left[2\,\pi\left((\bar{n}+1)(n+1)\right)^{1/2}\tau\right] \qquad 30$$

As is easily seen, the square root can be expanded as a constant plus a term linear in the difference between n and $\bar{n}$.

Note that some of the examples shown within Table 3 are repeats of examples in Table 2 for ease of review. Further, for the last example in the table, 50 TLMs are are. This last example was carried out to time $\tau$ = 20 $\tau_R$ and took 21 hours using parallel computation using 4 cores. The second special case will be considered next.

## All TLMs initially in the down state-Simulated Absorption

For this case, $r = \frac{N}{2}, m = -r, and\ P(r) = 1$. Starting with eq. (10) and applying the various restrictions one finds

$$\langle E^{-}E^{+}(t)\rangle = \left|\frac{\gamma}{\mu}\right|^{2} \{\bar{n} - S_4(\bar{n}, N, \gamma t)\}, \tag{30}$$

where

$$S_4(\bar{n}, N, \gamma t) = -4\sum_{n=0}^{\infty}\langle n|\rho_f(0)|n\rangle \sum_{j=0}^{Min[N,n]-1}\ \sum_{j'=j+1}^{Min[N,n]} (A^{*})_{n}^{\frac{N}{2},n-\frac{N}{2},j} A_{n}^{\frac{N}{2},n-\frac{N}{2},j'}\, Sin^2\left\{\frac{\left[q_{\frac{N}{2},\left(n-\frac{N}{2}\right),j} - q_{\frac{N}{2},\left(n-\frac{N}{2}\right),j'}\right]}{2}\Omega|\kappa|t\right\}$$
$$\times \sum_{p=0}^{\min[N,n]} pA_{n-p}^{\frac{N}{2},n-\frac{N}{2},j} (A^{*})_{n-p}^{\frac{N}{2},n-\frac{N}{2},j'} \tag{31}$$

This expression is very similar to eq. (14) but with some differences since the upper limits on the sums over j, j' and p depend on n and the eigenvectors $A_k^{r,c,j}$ are summed downward from the highest term rather than upwards from the lowest term. It can also be seen that for N TLMs, that there are N-1 unique terms, one each for every n term less than N. It is only for $n \geq N$ that all the terms look the same. Note 3: We can replace the Sin function in Eq. 31 to take advantage of time normalization., namely $Sin^2\left\{\pi\left[q_{\frac{N}{2},\left(n+\frac{N}{2}\right),j} - q_{\frac{N}{2},\left(n+\frac{N}{2}\right),j'}\right]\sqrt{\bar{n} + 1 + \Delta\tau}\right\}$.

### ***Numerical evaluation of $S_4(\bar{n}, N, \tau)$***

The calculations for this case appear to be no more difficult that for all the TLMs up, although the precision for which the calculations are performed must be raised to prevent round off errors or division by zero. As with the TLMs all up, all the photon distributions provided in Table 1 can be used for example calculations of $S_4(\bar{n}, N, \boldsymbol{\tau})$. Examples for all cases will not be provided but can be obtained using the supplied Mathematica notebook. Instead, the examples provided will at first focus on the use of the coherent distribution for various numbers of coherent photons and a comparison of the results for $S_4(\bar{n}, N, \boldsymbol{\tau})$ and $S_1(\bar{n}, N, \boldsymbol{\tau})$ (all TLMs up) for the same number of TLMs. Some interesting results will be seen. For a large number of coherent photons compared to the number of TLMs, the results will appear quite similar, although with differences. However, when the number of coherent photons approaches and then becomes smaller than the number of TLMs, striking differences between the 2 functions will occur. The results are presented in Table 4. There are nine sections in Table 4. Each section has both a left and right column and is 3 rows deep. The left and right columns are read vertically with the top row of each section the graphic for the photon density. The second row is the case for all TLMs up ($S_1(\bar{n}, N, \boldsymbol{\tau})$). The bottom row is for all TLMs down ($S_4(\bar{n}, N, \boldsymbol{\tau})$). Only resonant cases are provided although the reader may consider other cases using the supplied program. Section 1 contains the 2 cases of 10 coherent photons (Eq. 17 in Table 1) and 5 TLMs in the left column while the right contains the case of 100 coherent photons and 5 TLMs. As mentioned above, the case of mean photon number much larger than the TLM number results in very similar $S_1$ and $S_4$. In section 2 of the table, the cases of smaller mean photon number compared to the number of TLMs are presented. In fact, the left column has a mean photon number of 10 vs. 100 TLMs while the right column contains the case of 25 mean coherent photons vs. 100 TLMs. The cases for the TLMs initially up are much more chaotic than the case of all TLMs down. Again, the photon density was calculated from Eq. 17 in Table 1. The results in the left column for all TLMs down is very surprising in that a well-defined rise and fall in oscillations is seen. (Note: For the case of 10 coherent photons and 100 TLMs all down, we redefined the value of $\tau_R = 4\pi\sqrt{N - \bar{n} + \Delta}$ which results in the first peak at $\tau = 1$. )This behavior prompted a continued calculation in Section 3 of the table. Both left and right columns were for 100 TLMs. The left column was calculated for 1 coherent photon while the left was for 100 coherent photons and again the photon density is provided in Eq. 17, Table 1. It is again seen that there is a well-defined rise and fall of oscillation for the left column with all TLMs down. That is not the case for the right column but another unexpected result is seen. Namely that the average number of absorbed photons is more than have the number of mean photons. This is surprising since in all previous calculations for TLMs in the up state, the maximum number of mean emitted photons was never more than ½ the number of TLMs and often much smaller (the larger the number of mean photons, the closer to ½ the mean number of emitted photons for the resonant case). In the case here, the mean number of absorbed photons is approaching 60. (Admittedly, this number is now dependent on photon number rather than TLM number). We will discuss this somewhat later.

Based on the previous results, it was decided to continue the investigation of a small mean number of photons vs. the number of TLMs but with the other photon densities provided in Table 1. Section 4 of the table contains the results for the Thermal photon density in Eq. 18 was used for both left and right columns. The mean photon number was 1 for the left column while the right had a mean photon number of 10. A TLM number of 100 was again used. Again, we see a very surprising results, namely the expected chaotic behavior for the TLMs up but a well-defined oscillation for the TLMs down. This was not expected since all previous calculations with the Thermal Photon density always displayed chaotic behavior unless strong non-resonance was applied.

Sections 5, 6, 7, 8 and 9 of the table contain similar examples for the other photon densities in Table 1. Note: Section 6 of the table is interesting since the photon density has non-zero values for every other value of n. For the down case this results in echoes at half integer values of normalized time. For all the remaining cases the number of TLMs was constrained to 50 to limit the time it took to generate the figures . Section 5 of the table has the results for the photon densities defined by Mixed Coherent and Thermal State (Eq. 20) in the left column and the Squeezed Coherent State (Eq. 24) in the right. Section 6 of the table has the results for the photon densities defined by Squeezed Thermal State (Eq. 23) in the left column and Squeezed Vacuum State (Eq. 21) in the right. Section 7 of the table has the results for the Squeezed Fock State (Eq. 22) in the left column and the DSTS (Eq. 26) in the right. Section 8 of the table has the results for the Mixed Squeezed Coherent plus Thermal Noise State (Eq. 25) in the left column and the Displaced Number State (Eq. 27) in the right. The final section 9 of the table, contains only the Squeezed Displaced Number State photon density (Eq. 28). In nearly all of these cases, the time development of $S_1$ is rather chaotic although if a few cases there is some coherence for a short time. However, again for all TLMs down we see rather a well-defined rise and fall of oscillation except in those cases with large values of variance in the photon number such as the left column of section 6 and the left column of section 7.

# Discussion

In this article, simulated emission and absorption of radiation in a single mode (N-TLMs) for various photon distributions has been examined. The original purpose of this paper was to complete (wrap up) the results from the previous work presented in references [5, 7, 10 and 15]. However, as seen below and unexpected results was encountered.

The examples of simulated emission have produced the expected results based on previous work for a single TLM for the various photon density distributions and the results for the coherent photon density distribution represented by Eq. 17 for various TLM values. On the other hand, the results for simulated absorption, especially for the case of small mean photon number compared to the number of TLMs was unexpected. (Namely the well-defined rise and fall of oscillation as a function of time. This was unexpected since values of photon number approaching or slightly larger than the number of TLMs is considerably more chaotic. This may be due to either a coherent effect of the large number of TLMs all at equivalent mode positions or due to the photons only interacting with a small number of the TLMs. This last conjecture is postulated due to the chaotic effect seen for small photon number when the photon variance is large as seen in sections 6 and 7 of Table 4.

In the introduction, we noted that we used time normalization $\tau_{\mathrm{R}} = 2\pi\frac{\sqrt{\bar{\mathrm{n}}+1+\Delta}}{\gamma}$ for TLMs up. In addition, we used time normalization $\tau_R = \frac{4\pi\sqrt{N-\bar{n}+\Delta}}{\gamma}$ for TLMs down. For almost all case when the time evolution of $S_1$ $or$ $S_4$ wasn't chaotic, photon echo revivals occurred at integer values of normalized time. Exceptions occurred when the photon density was non-zero for every other value of n. Then revivals were seen at half integer values of normalized time. (See Table 2 for the squeezed vacuum state and the squeezed Fock state.) In one case (the first entry in Table 2), a very small revival is seen at the first half integer in normalized time. It should be noted that the value of $\tau_R$ is only approximate in some cases.

As mentioned above, there is an additional interesting observation concerning simulated absorption which occurs when the mean number of photons approaches the number of TLMs. Namely the mean number of photons absorbed becomes larger than ½ the number of the initial mean number. This behavior was observed for the coherent case [Eq. 17] for all the number TLMs considered above 5 TLMs and when fit to a second order equation (the second order component has a very small coefficient) showed remarkably small variance from the fit (See Fig. 1 below).(See Table 3 for the Coherent case with an initial photon number of 100 and the TLM number of 100.)

Note

There is one included file with this submission. There is a Wolfram Mathematica notebook (version 12.1.1.0) used to provide the calculations and figures presented in this paper. Normalized time was included in this notebook. The previous version contained two ancillary files.

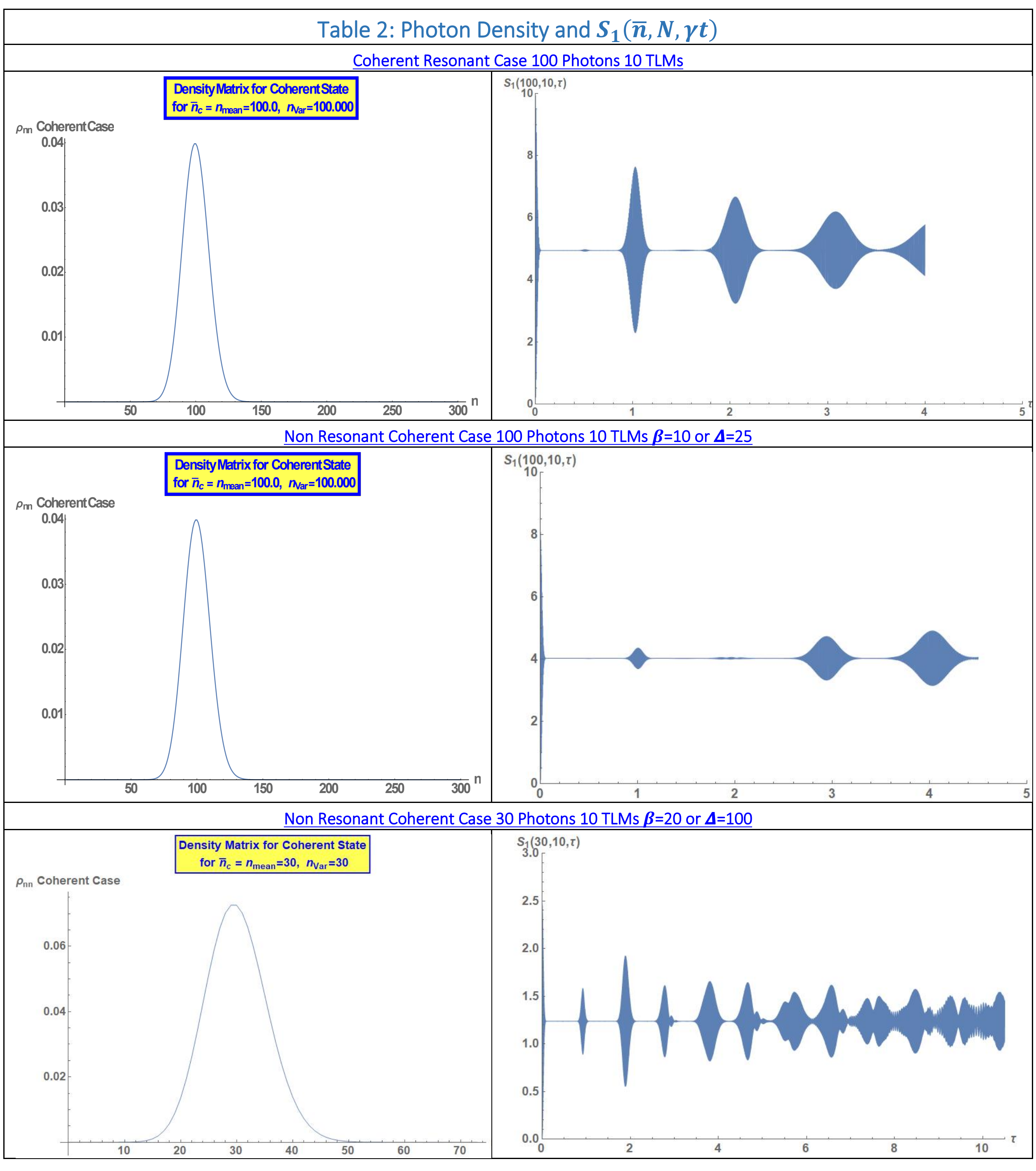

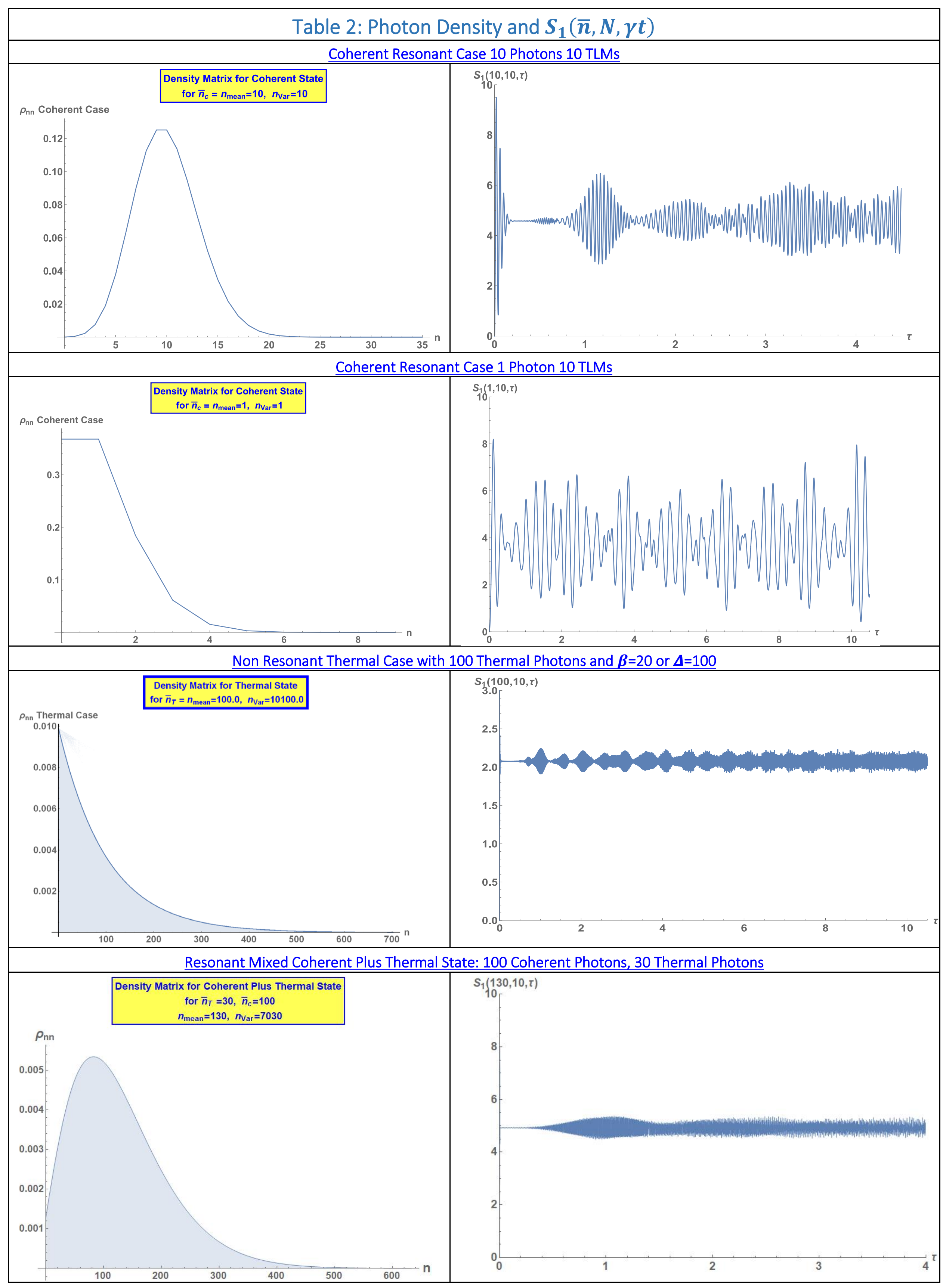

## Table 2: Photon Density and $S_1(\bar{n}, N, \gamma t)$

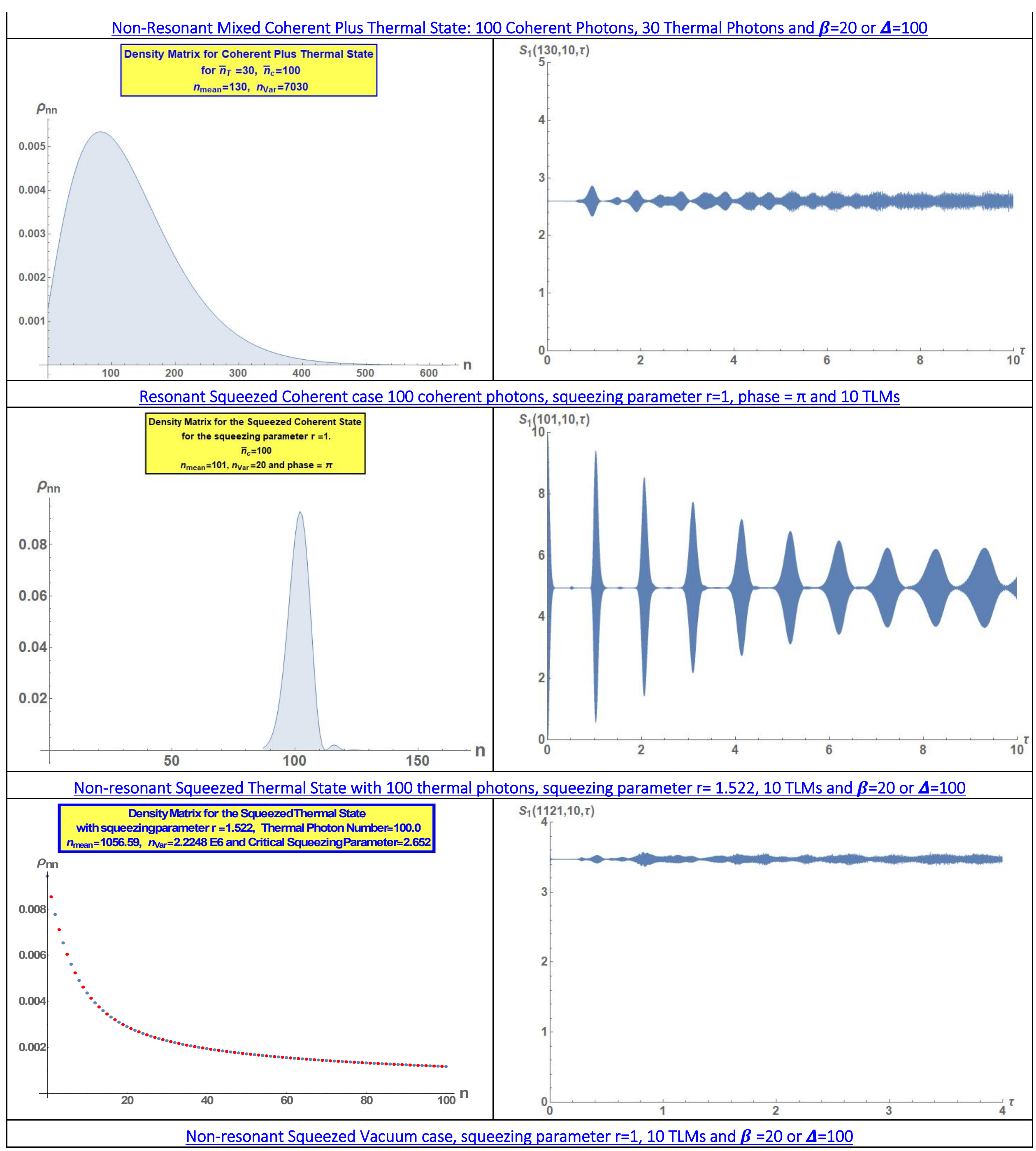

## Table 2: Photon Density and $S_1(\overline{n}, N, \gamma t)$

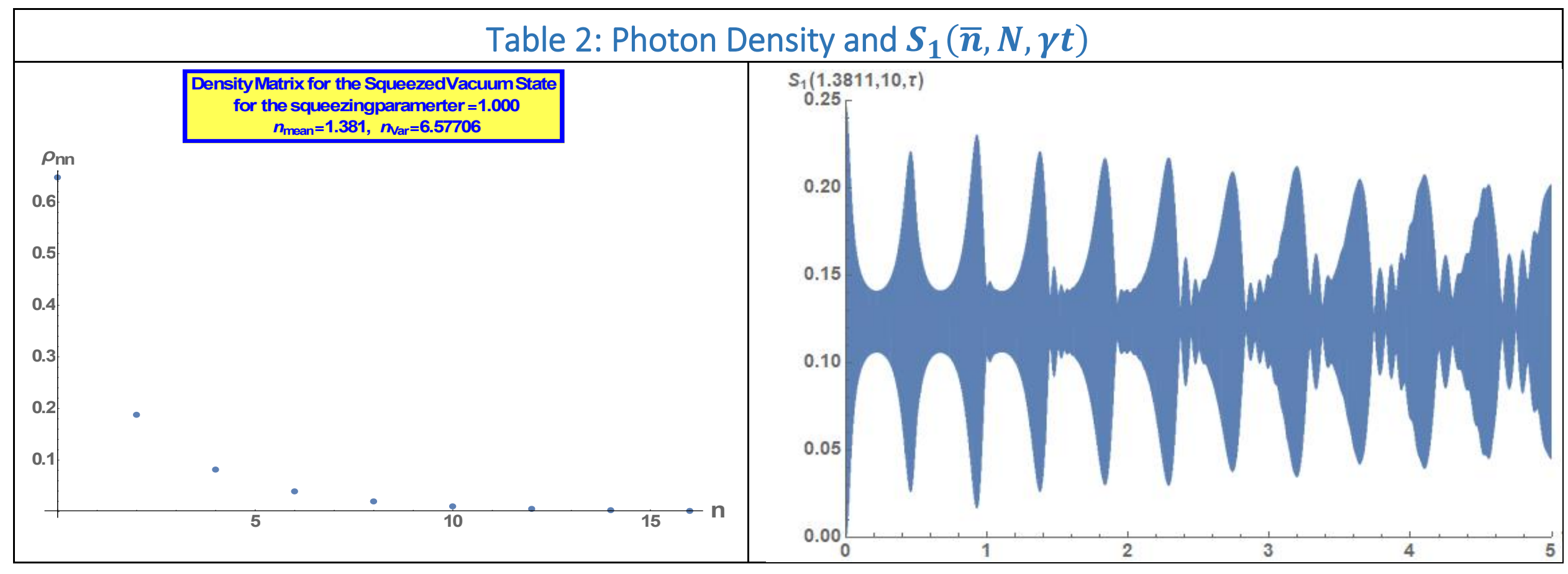

Resonant Squeezed Fock State with 100 Initial Photons, squeezing parameter r=1 and 10 TLMs

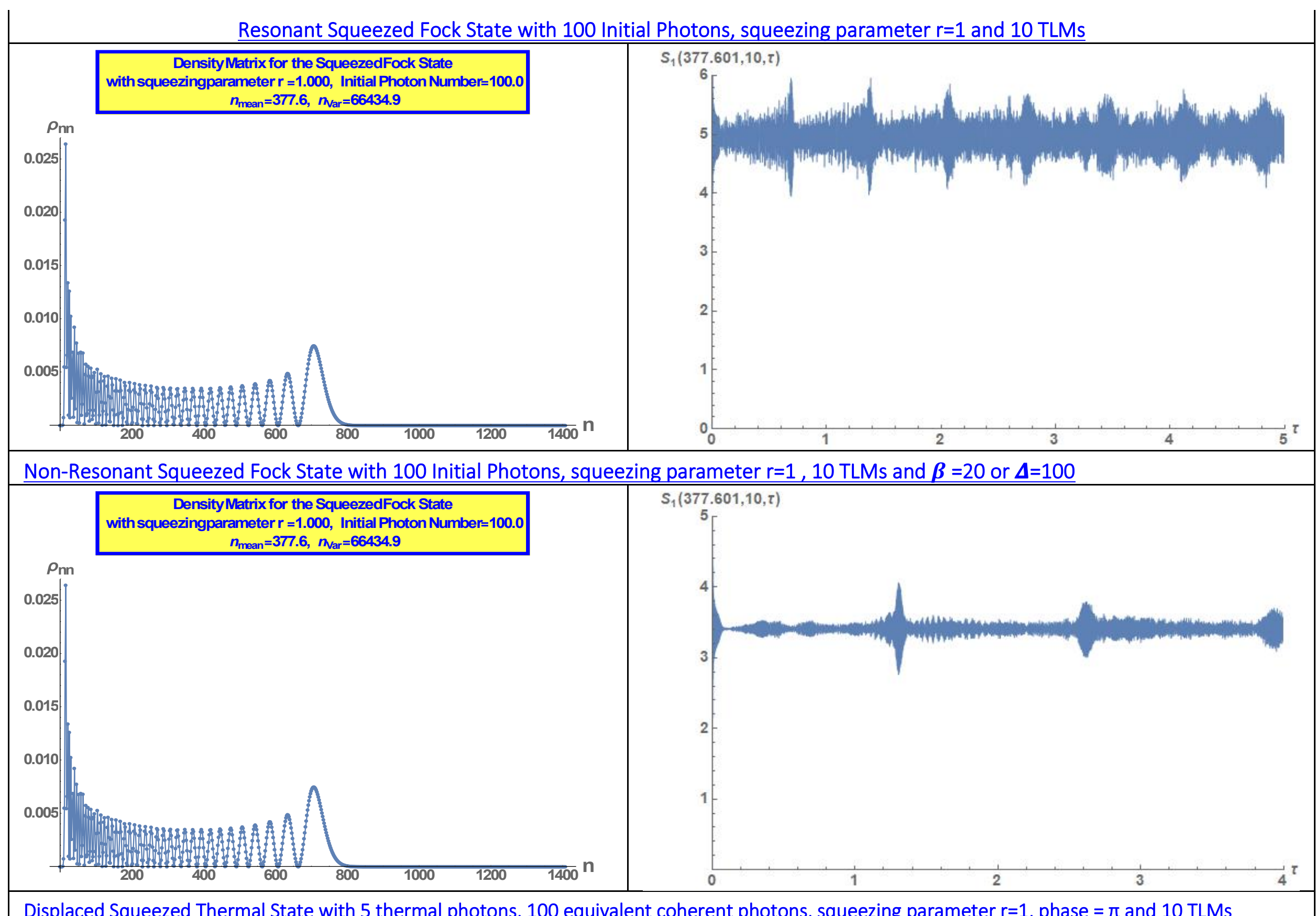

Non-Resonant Squeezed Fock State with 100 Initial Photons, squeezing parameter r=1 , 10 TLMs and $\boldsymbol{\beta}$ =20 or $\boldsymbol{\Delta}$=100

Displaced Squeezed Thermal State with 5 thermal photons, 100 equivalent coherent photons, squeezing parameter r=1, phase = π and 10 TLMs

## Table 2: Photon Density and $S_1(\bar{n}, N, \gamma t)$

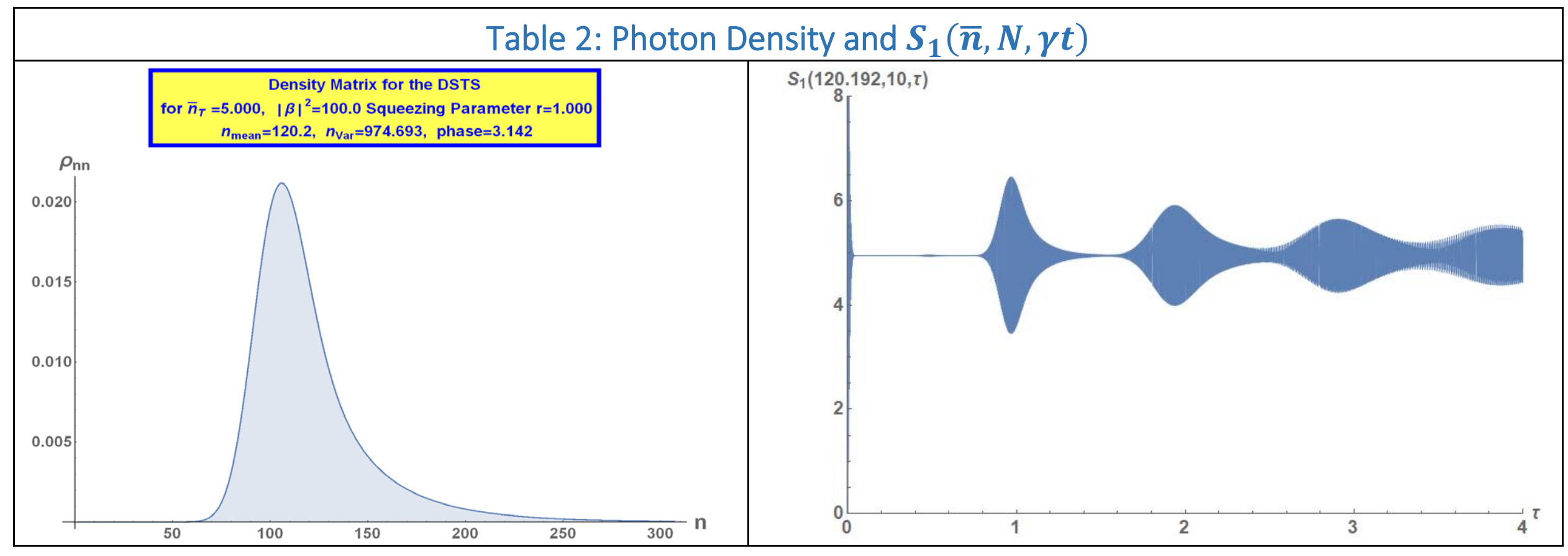

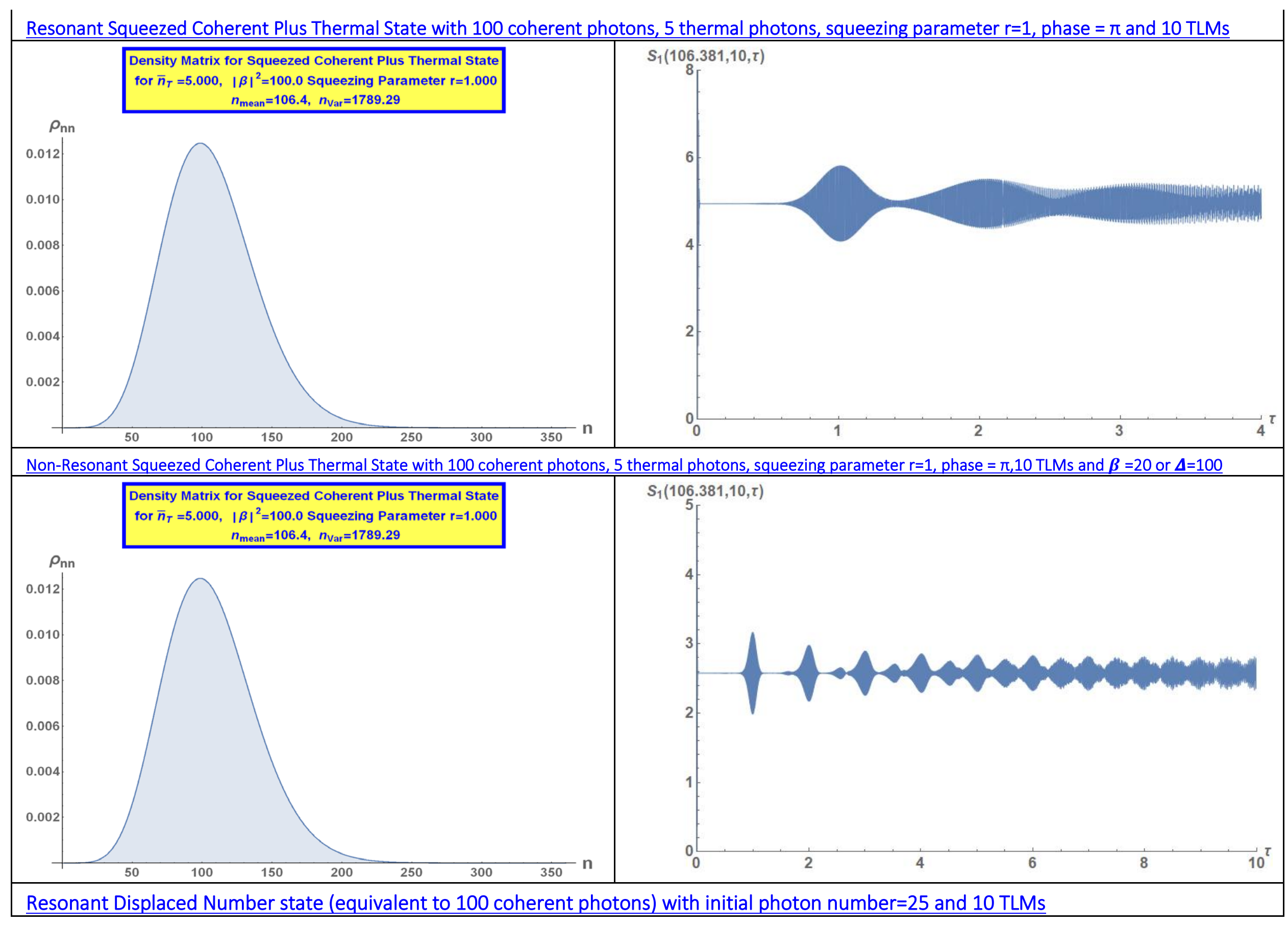

Resonant Squeezed Coherent Plus Thermal State with 100 coherent photons, 5 thermal photons, squeezing parameter r=1, phase = π and 10 TLMs

Non-Resonant Squeezed Coherent Plus Thermal State with 100 coherent photons, 5 thermal photons, squeezing parameter r=1, phase = π,10 TLMs and $\boldsymbol{\beta}$ =20 or $\boldsymbol{\Delta}$=100

Resonant Displaced Number state (equivalent to 100 coherent photons) with initial photon number=25 and 10 TLMs

Table 2: Photon Density and $S_1(\bar{n}, N, \gamma t)$

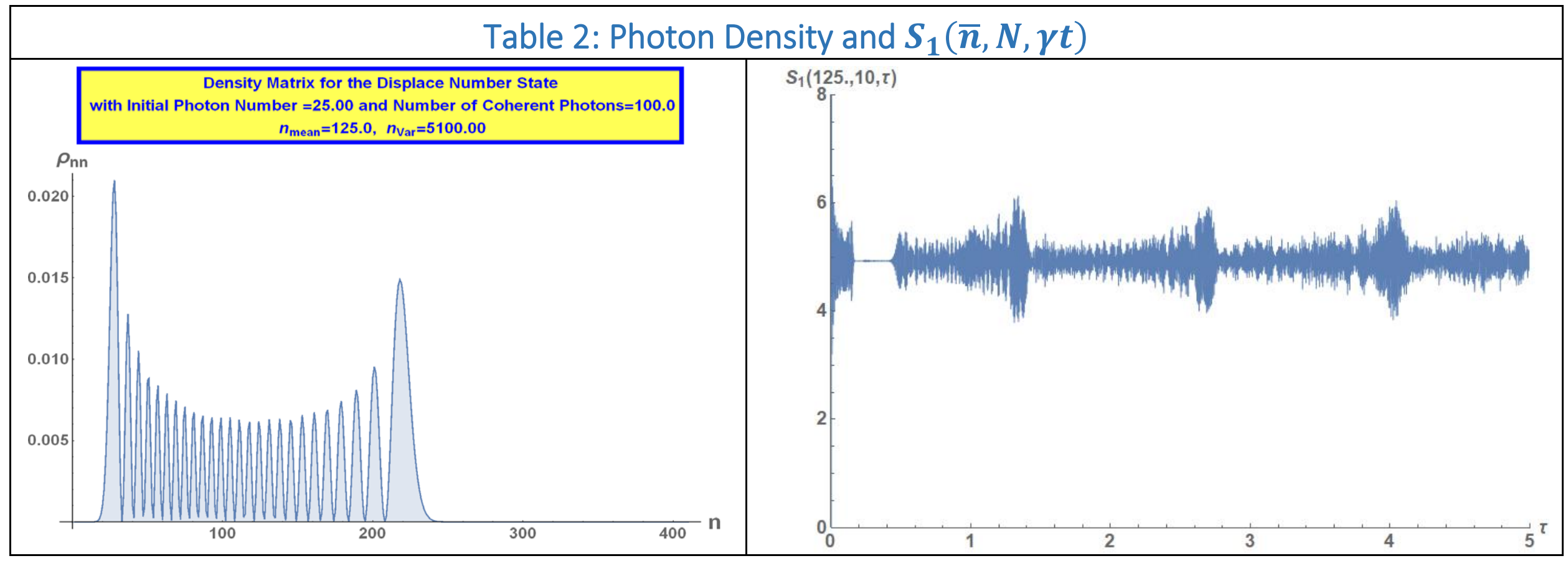

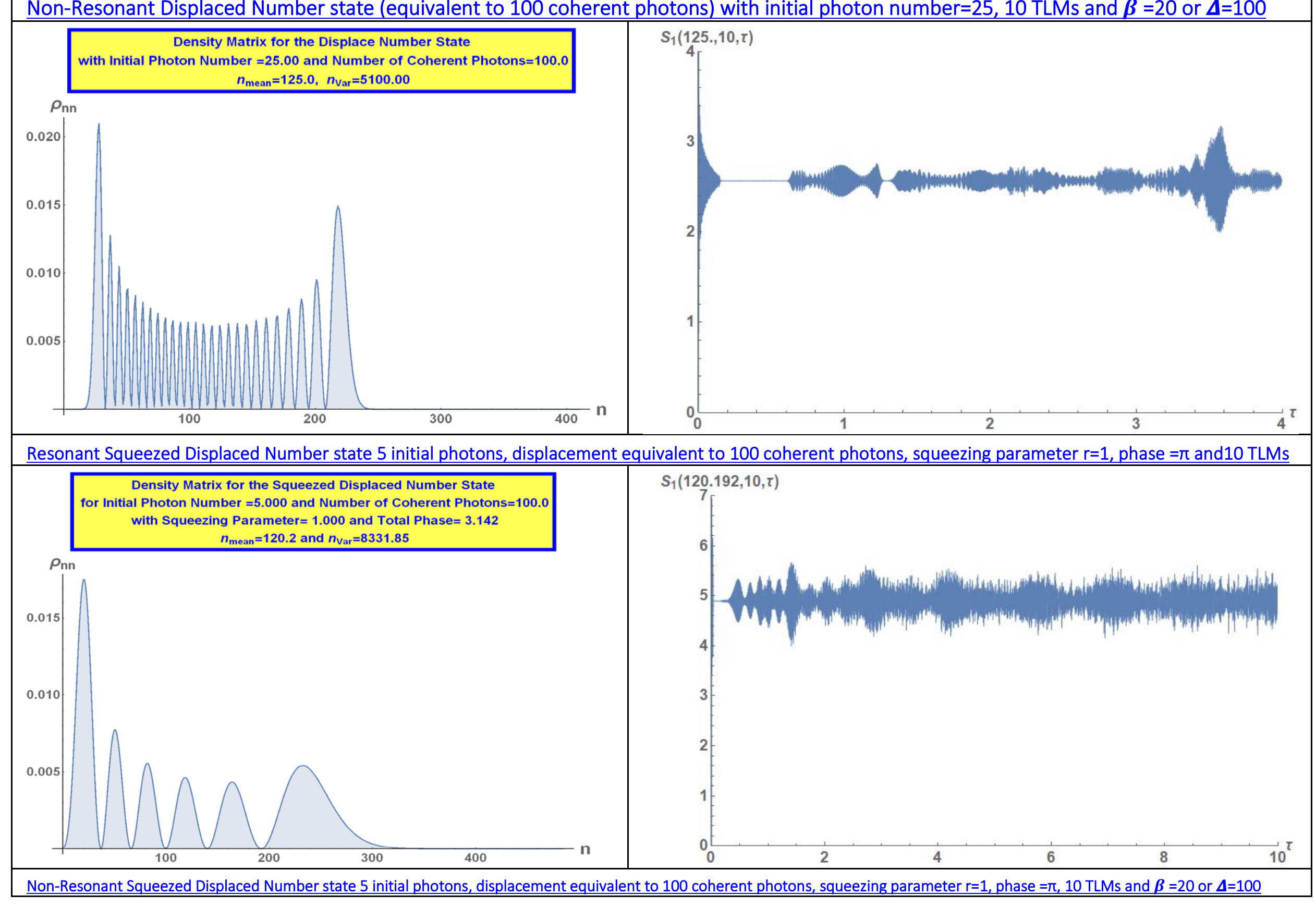

Non-Resonant Displaced Number state (equivalent to 100 coherent photons) with initial photon number=25, 10 TLMs and $\boldsymbol{\beta}$ =20 or $\boldsymbol{\Delta}$=100

Resonant Squeezed Displaced Number state 5 initial photons, displacement equivalent to 100 coherent photons, squeezing parameter r=1, phase =π and10 TLMs

Non-Resonant Squeezed Displaced Number state 5 initial photons, displacement equivalent to 100 coherent photons, squeezing parameter r=1, phase =π, 10 TLMs and $\boldsymbol{\beta}$ =20 or $\boldsymbol{\Delta}$=100

Table 2: Photon Density and $S_1(\bar{n}, N, \gamma t)$

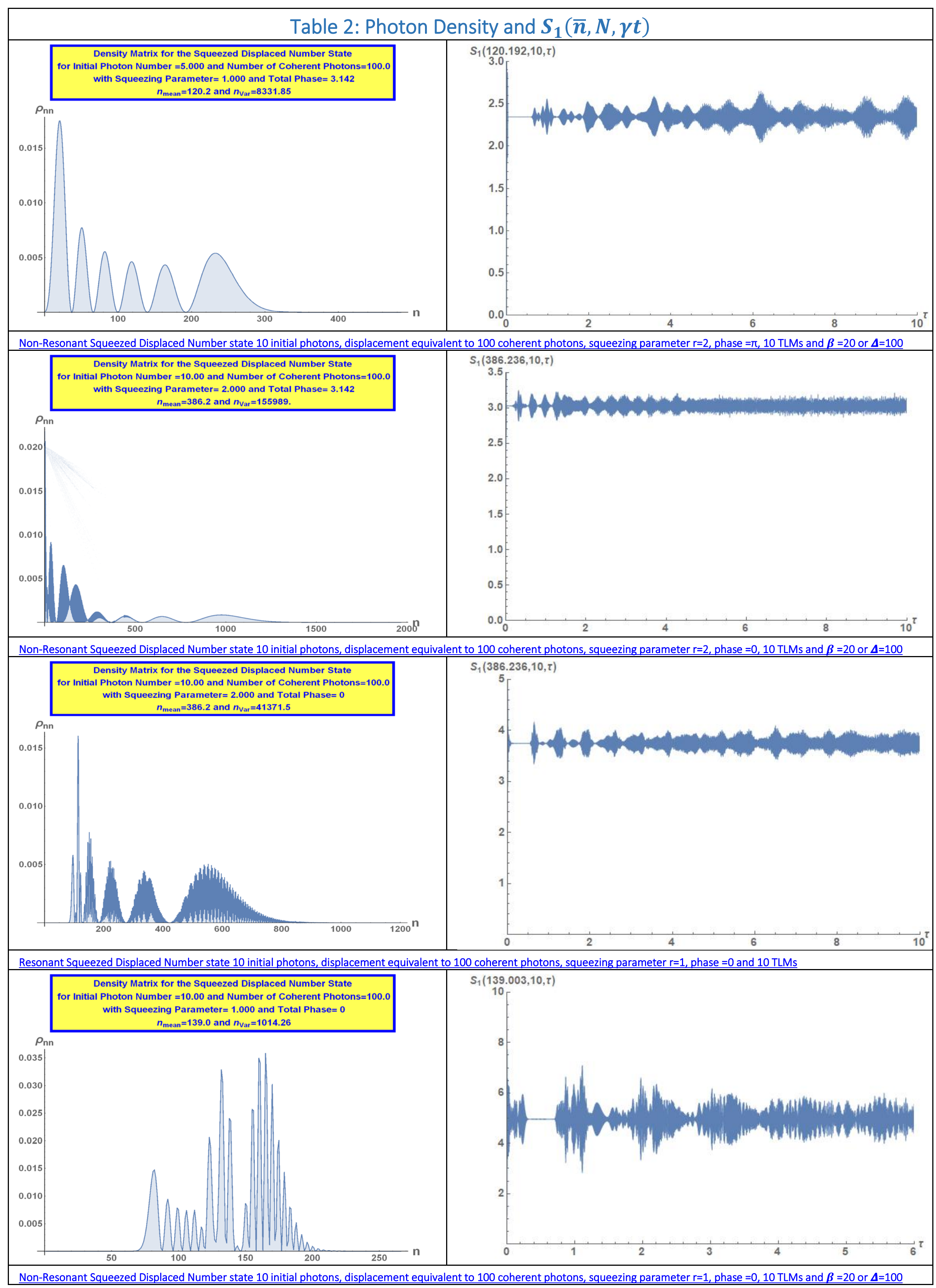

Non-Resonant Squeezed Displaced Number state 10 initial photons, displacement equivalent to 100 coherent photons, squeezing parameter r=2, phase =π, 10 TLMs and $\beta$ =20 or $\Delta$=100

Non-Resonant Squeezed Displaced Number state 10 initial photons, displacement equivalent to 100 coherent photons, squeezing parameter r=2, phase =0, 10 TLMs and $\beta$ =20 or $\Delta$=100

Resonant Squeezed Displaced Number state 10 initial photons, displacement equivalent to 100 coherent photons, squeezing parameter r=1, phase =0 and 10 TLMs

Non-Resonant Squeezed Displaced Number state 10 initial photons, displacement equivalent to 100 coherent photons, squeezing parameter r=1, phase =0, 10 TLMs and $\beta$ =20 or $\Delta$=100

Table 2: Photon Density and $S_1(\overline{n}, N, \gamma t)$

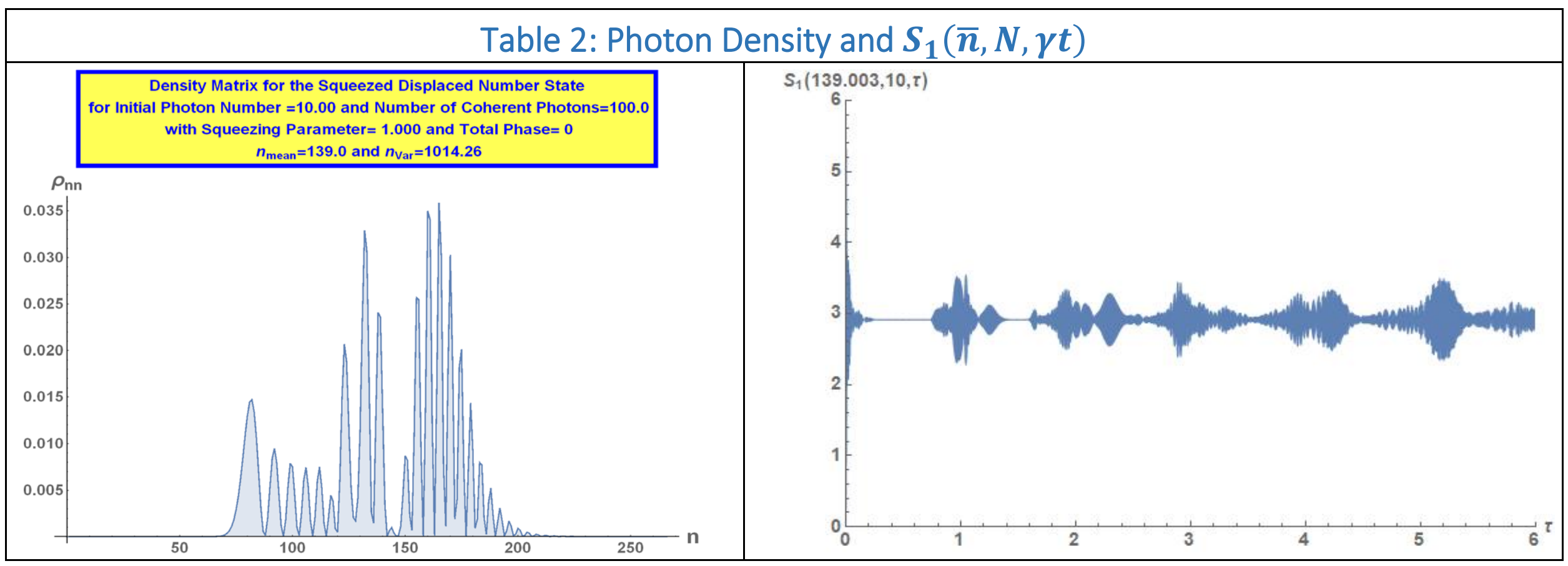

Table 3: Effect of Increasing Photon number on $S_1(\bar{n}, N, \gamma t)$ *for the coherent photon density*

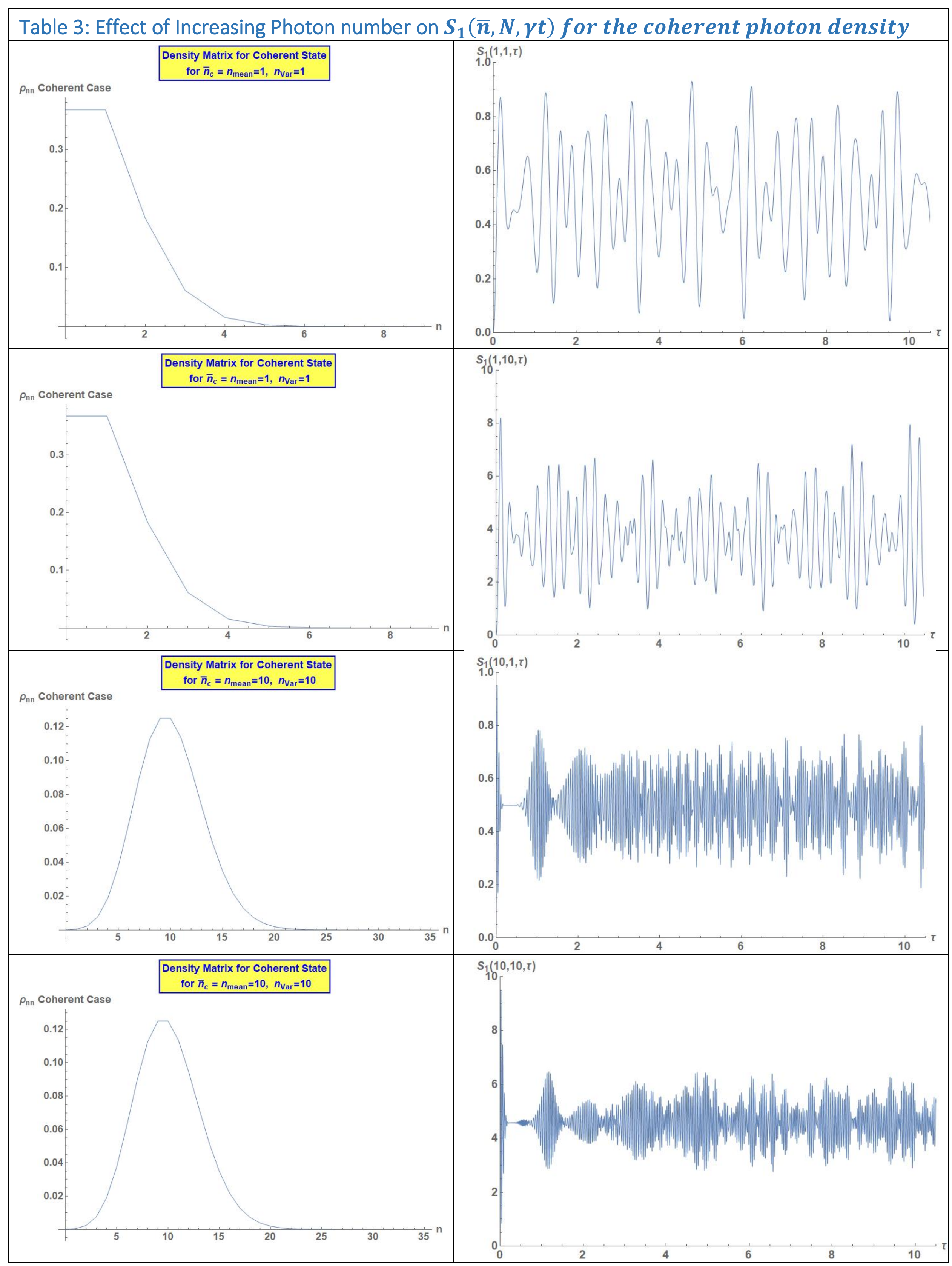

Table 3: Effect of Increasing Photon number on $S_1(\overline{n}, N, \gamma t)$ *for the coherent photon density*

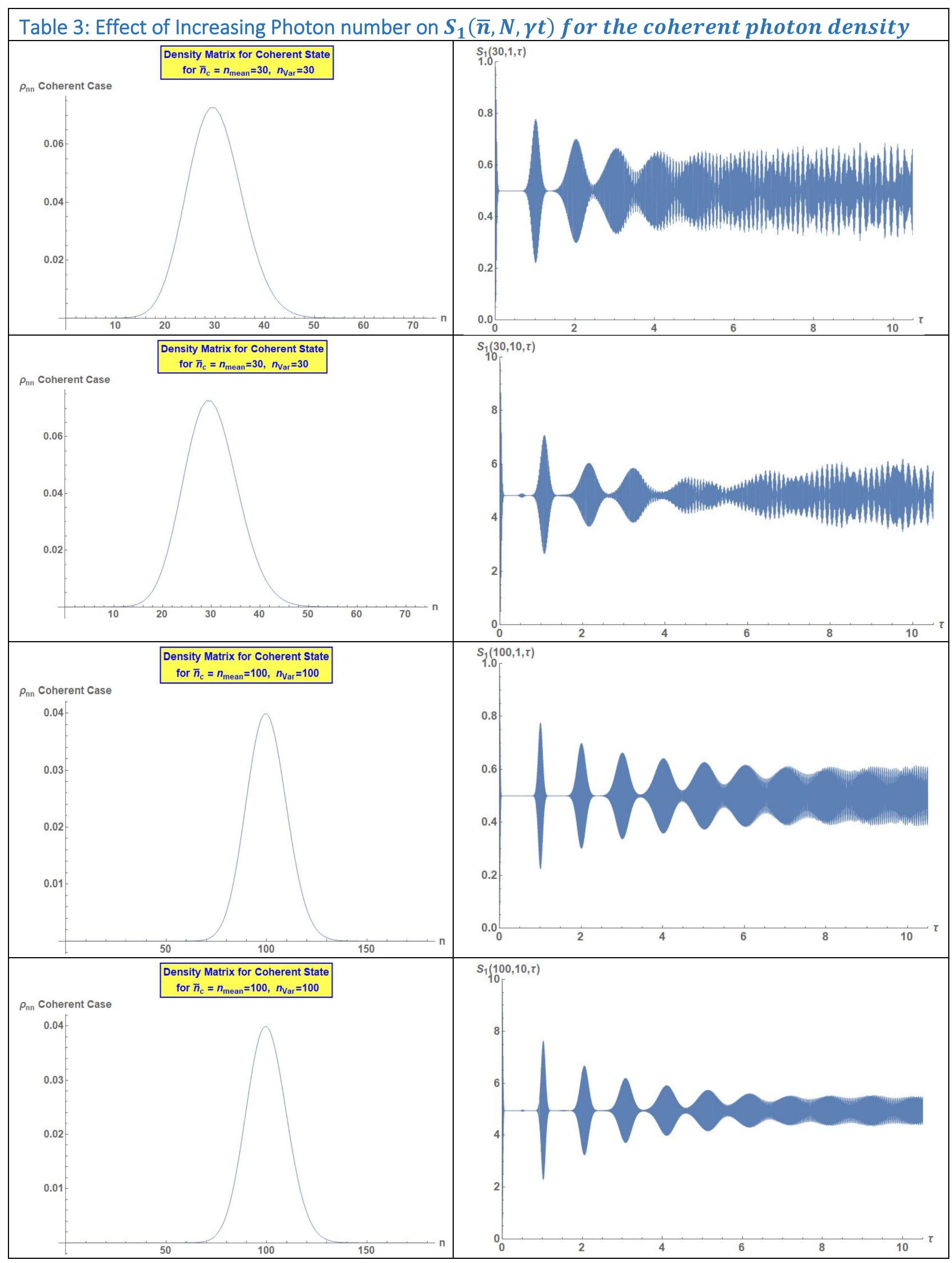

Table 3: Effect of Increasing Photon number on $S_1(\bar{n}, N, \gamma t)$ *for the coherent photon density*

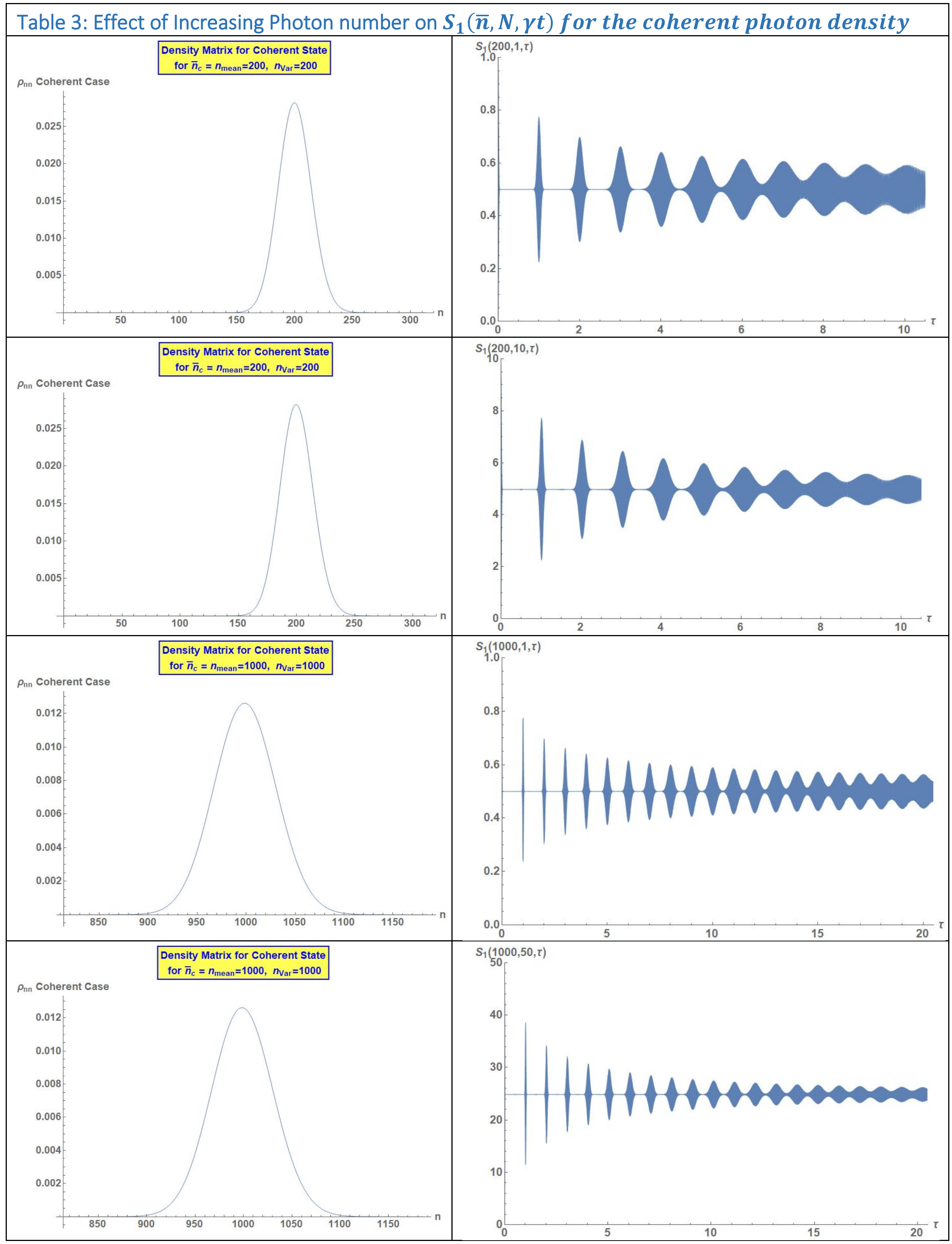

Table 4: Photon Density with a comparison between $\mathbf{S_1(\bar{n}, N, \tau)}$ and $\mathbf{S_4(\bar{n}, N, \tau)}$

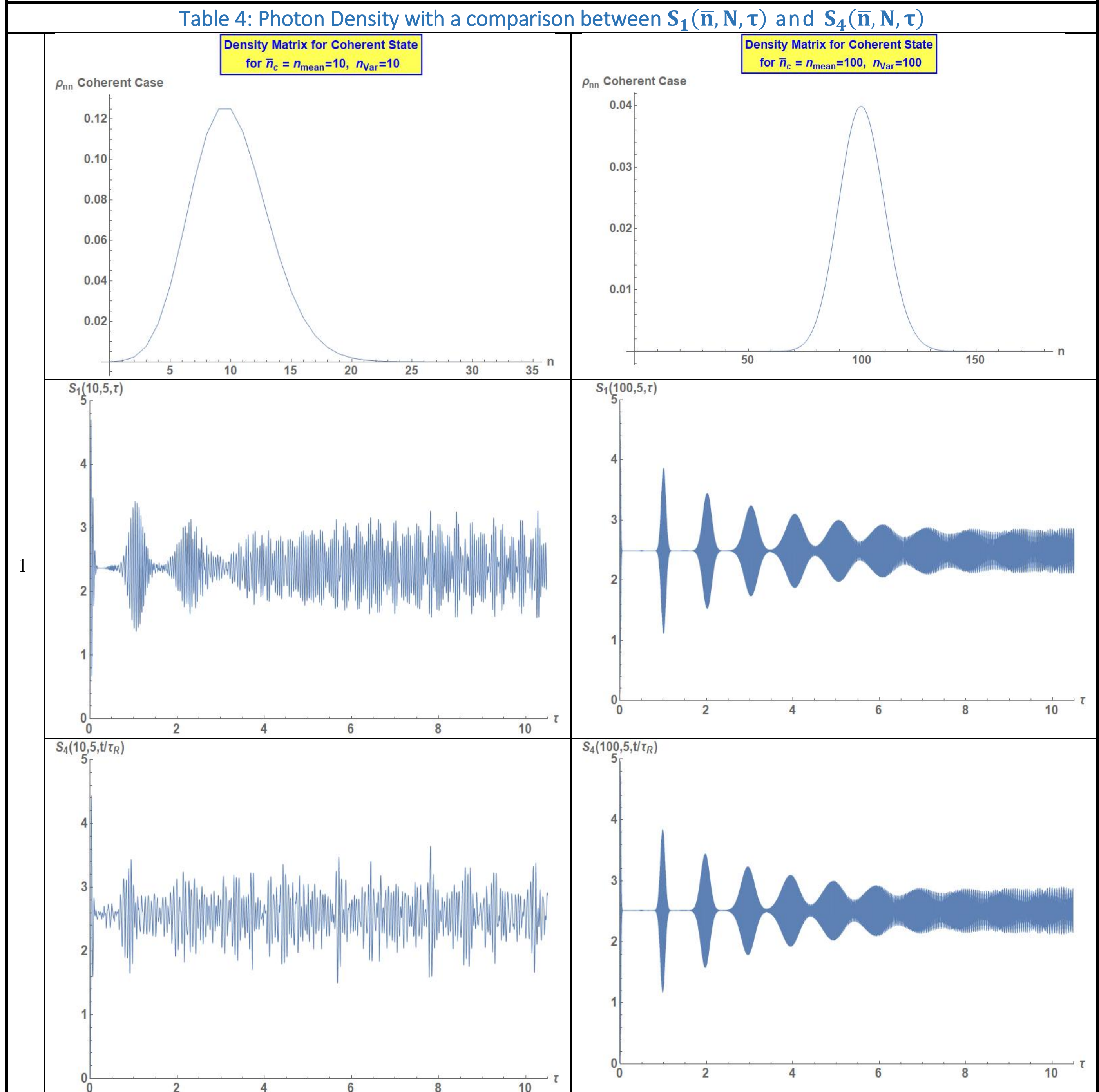

Table 4: Photon Density with a comparison between $\mathbf{S_1(\bar{n}, N, \tau)}$ and $\mathbf{S_4(\bar{n}, N, \tau)}$

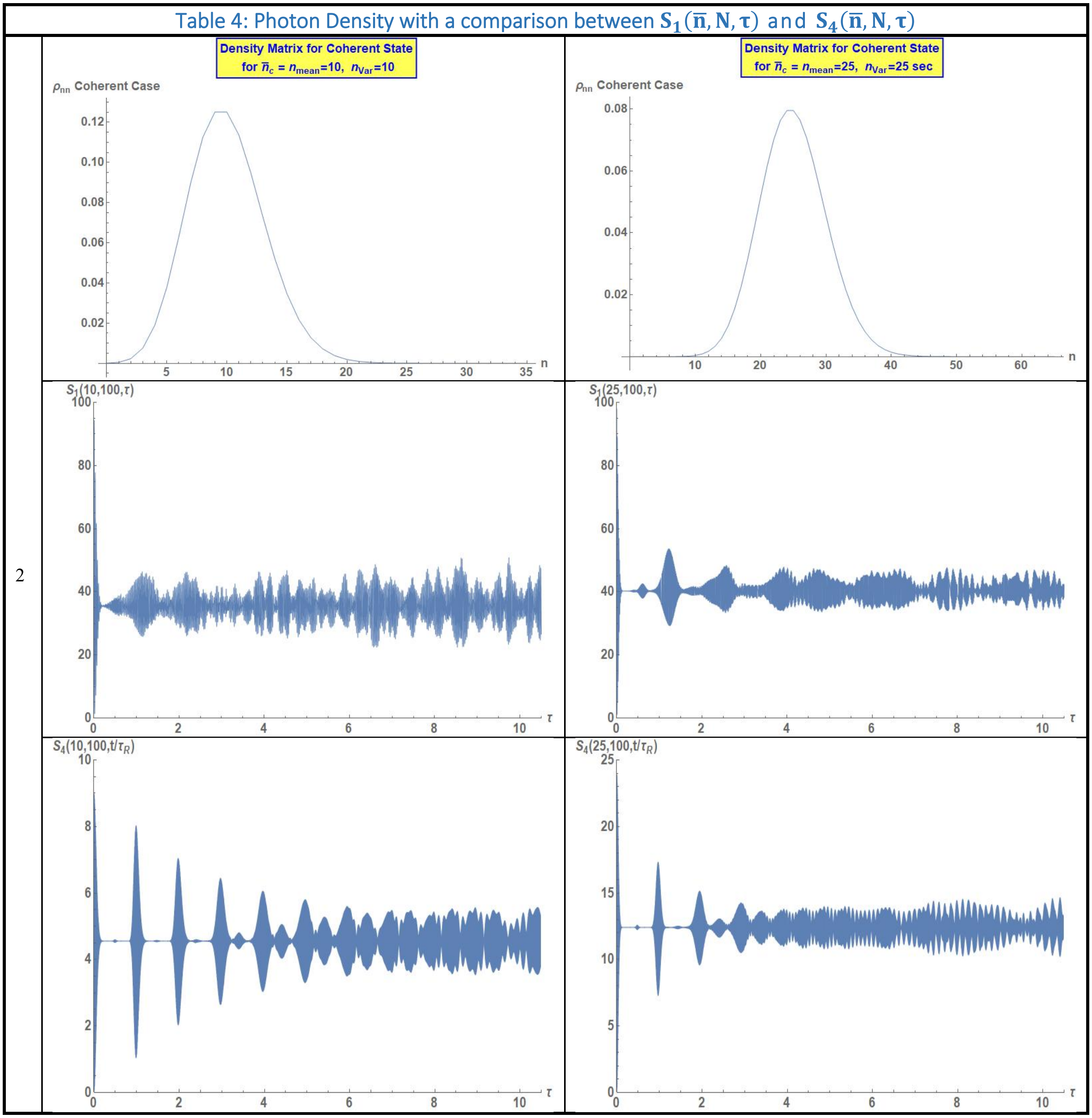

Table 4: Photon Density with a comparison between $\mathbf{S_1(\bar{n}, N, \tau)}$ and $\mathbf{S_4(\bar{n}, N, \tau)}$

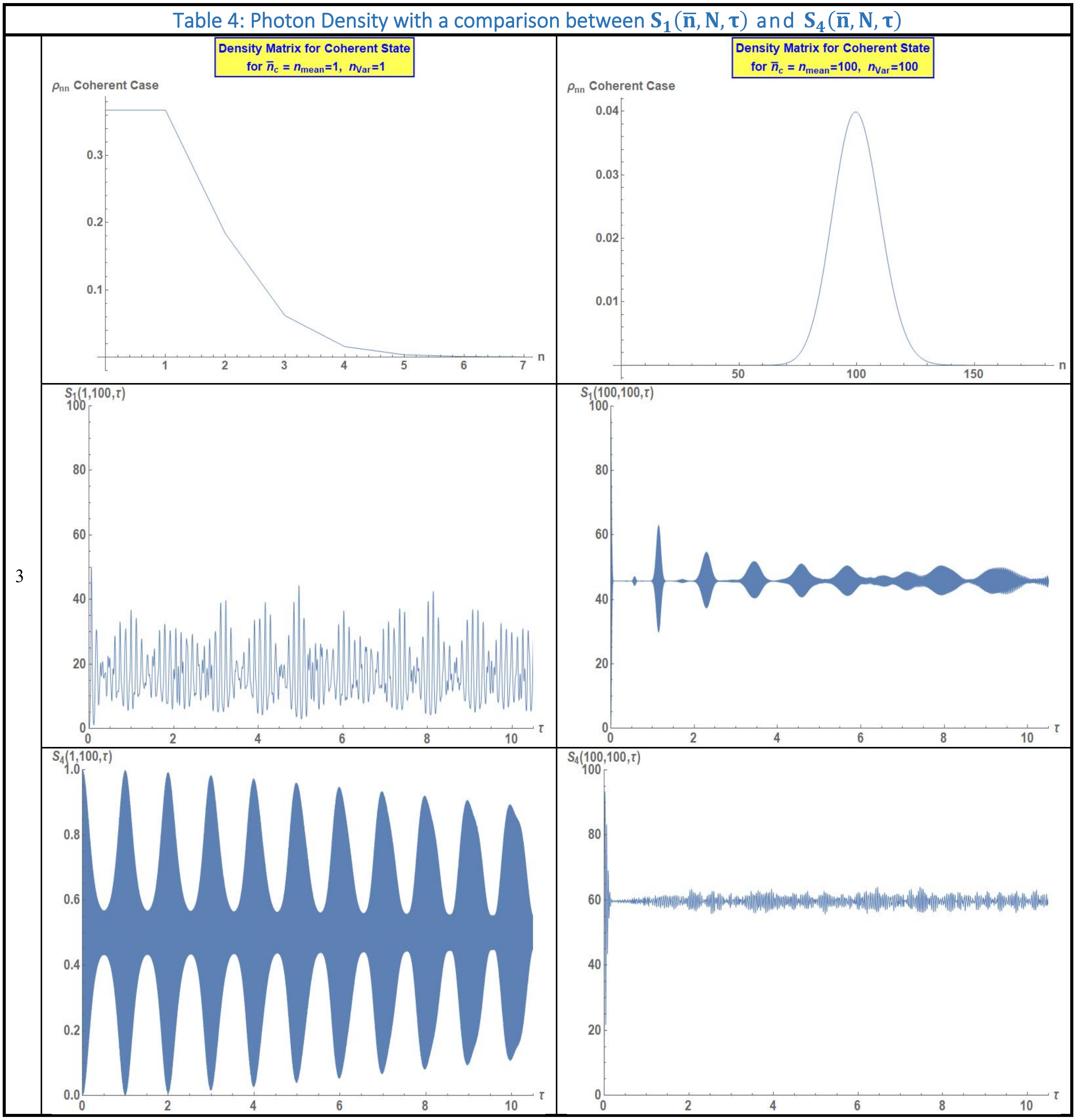

Table 4: Photon Density with a comparison between $\mathbf{S_1(\bar{n}, N, \tau)}$ and $\mathbf{S_4(\bar{n}, N, \tau)}$

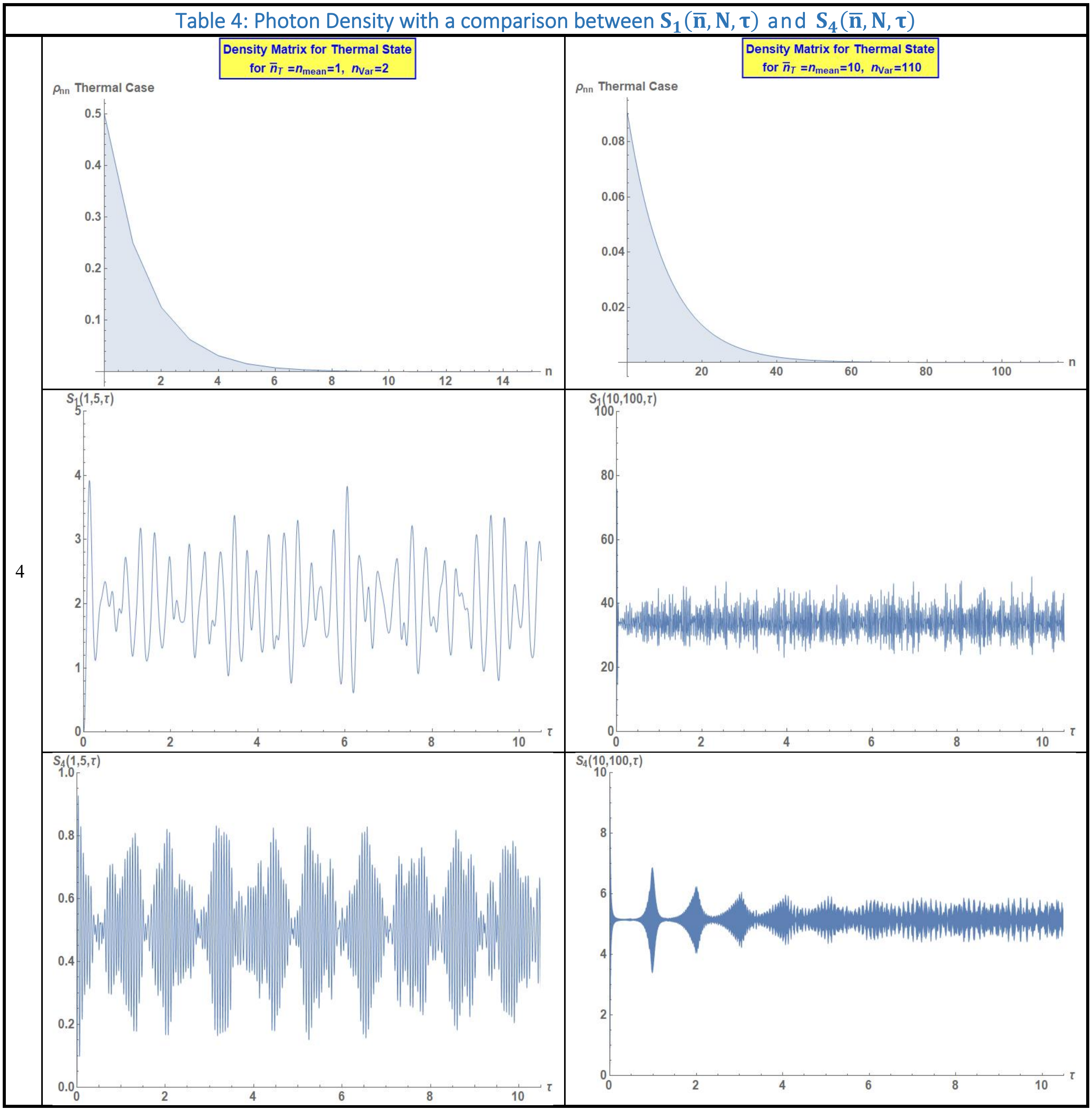

Table 4: Photon Density with a comparison between $\mathbf{S_1(\bar{n}, N, \tau)}$ and $\mathbf{S_4(\bar{n}, N, \tau)}$

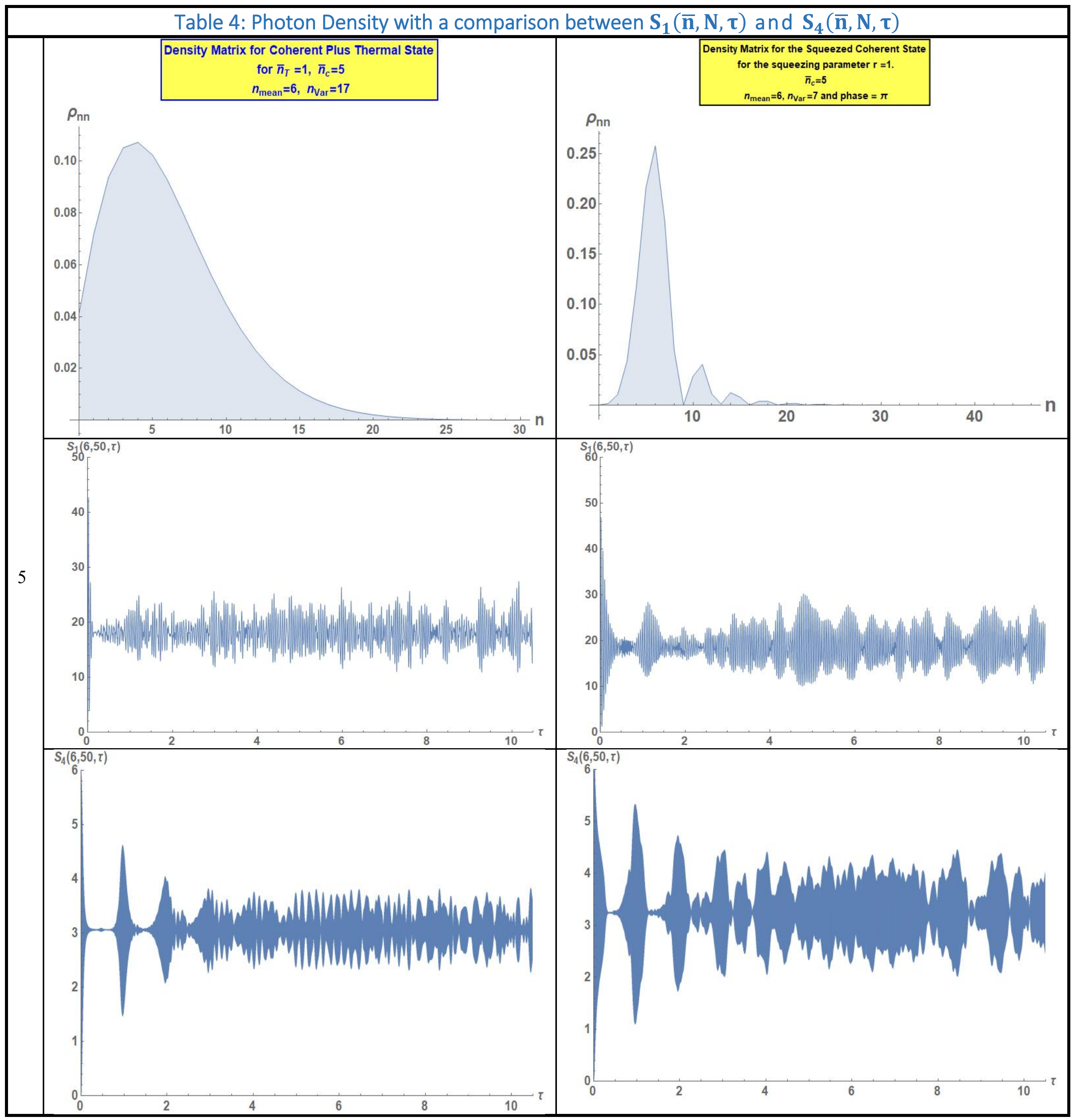

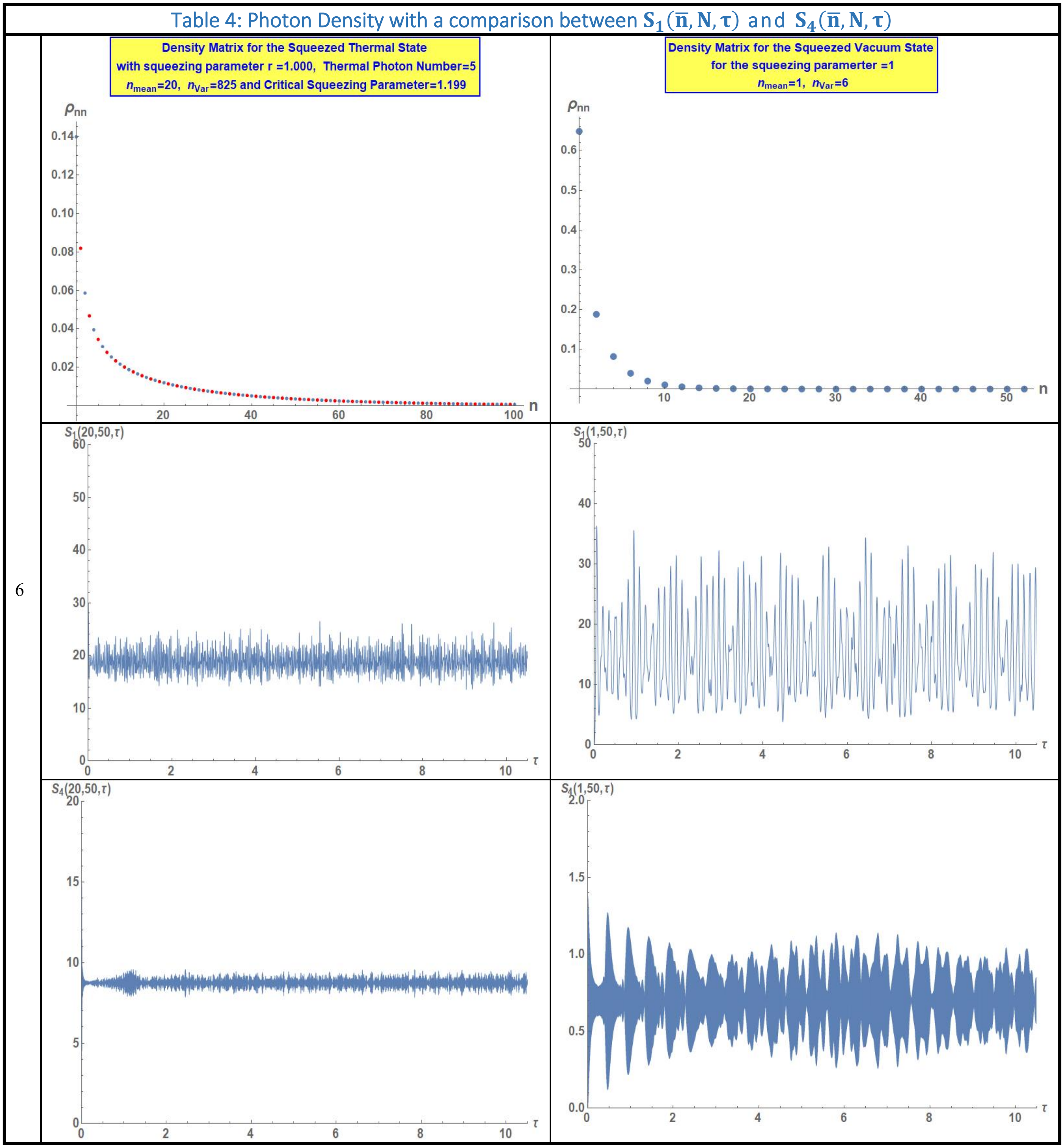

Table 4: Photon Density with a comparison between $\mathbf{S_1(\bar{n}, N, \tau)}$ and $\mathbf{S_4(\bar{n}, N, \tau)}$

Table 4: Photon Density with a comparison between $\mathbf{S_1(\bar{n}, N, \tau)}$ and $\mathbf{S_4(\bar{n}, N, \tau)}$

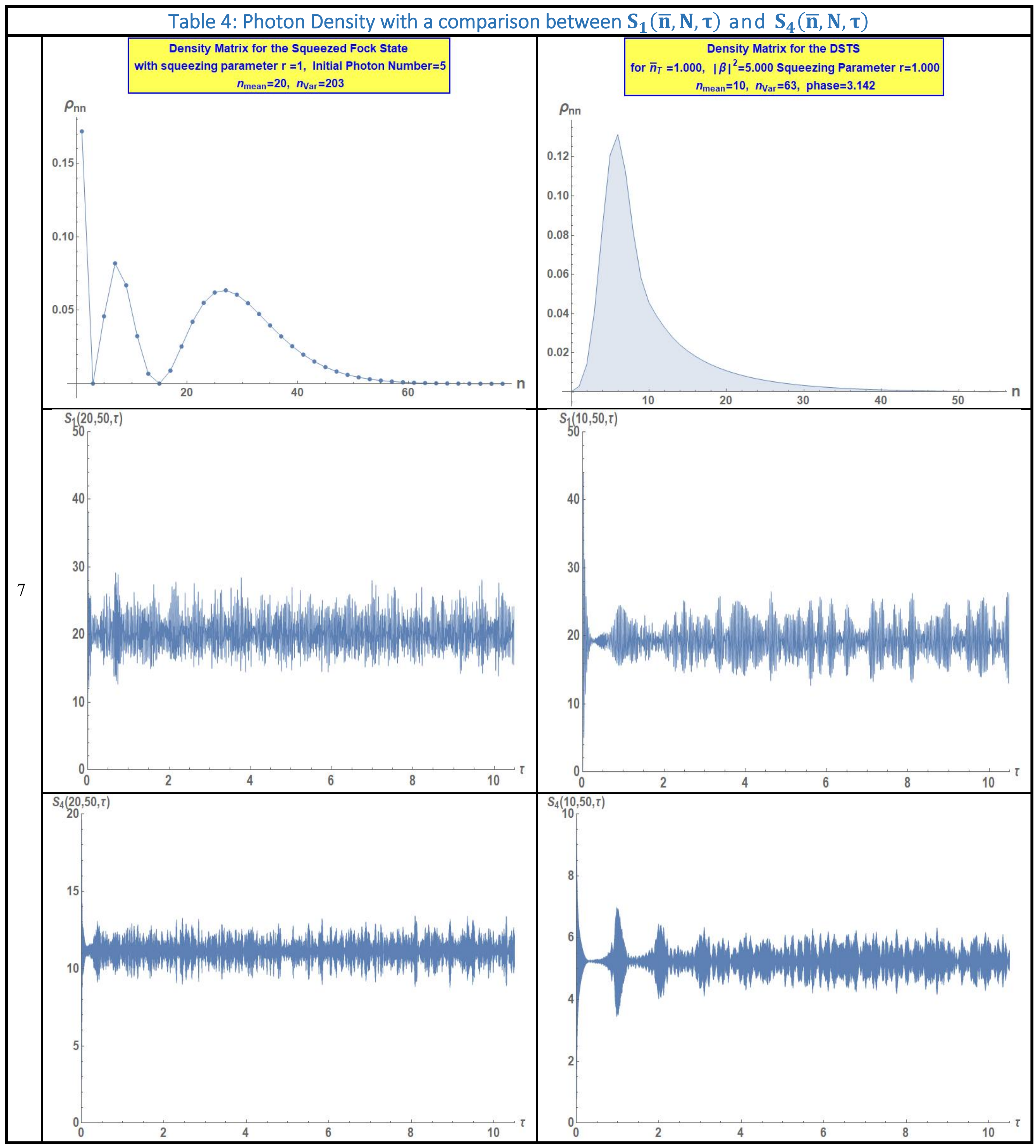

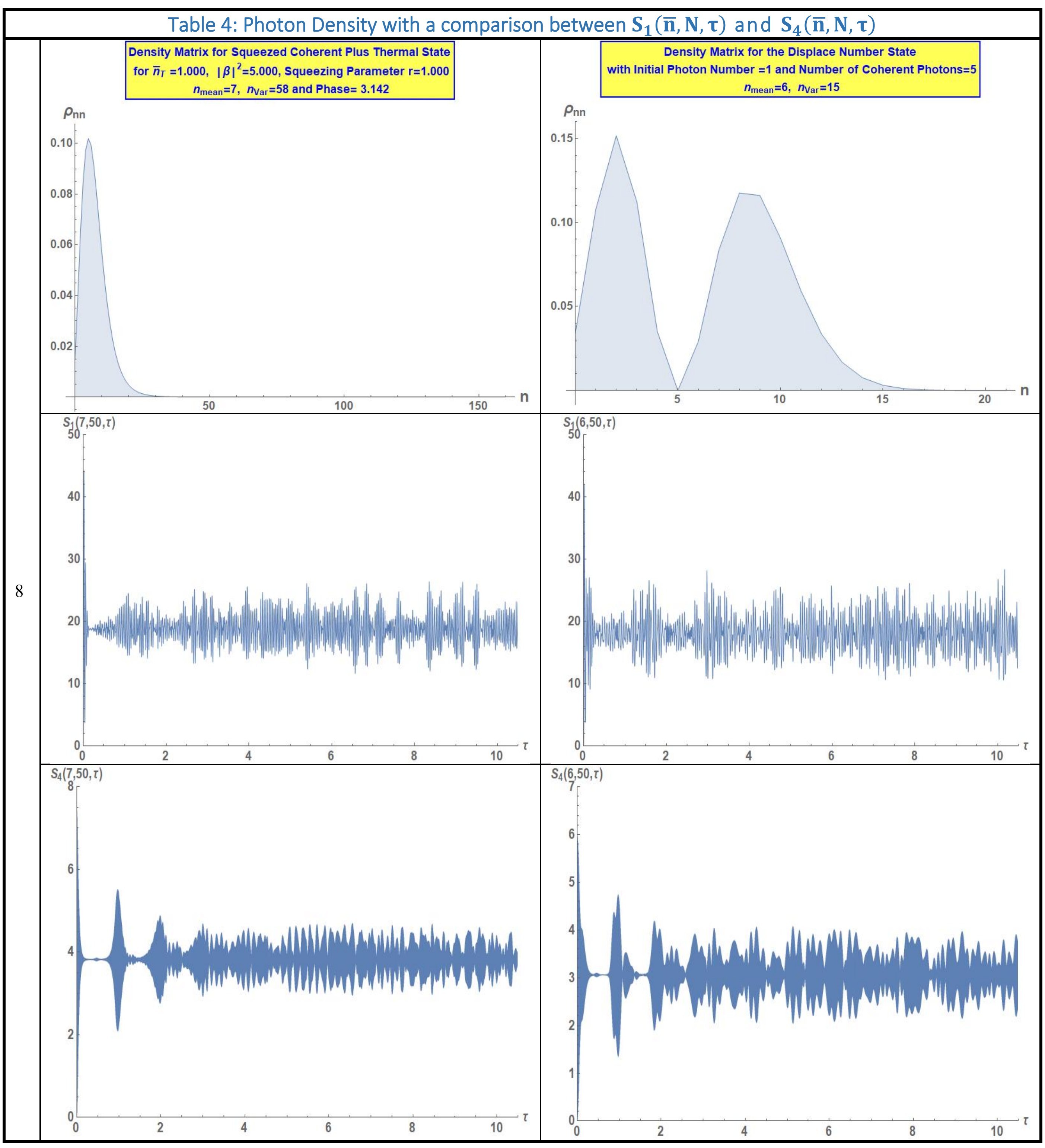

Table 4: Photon Density with a comparison between $\mathbf{S_1(\bar{n}, N, \tau)}$ and $\mathbf{S_4(\bar{n}, N, \tau)}$

Table 4: Photon Density with a comparison between $\mathbf{S_1(\bar{n}, N, \tau)}$ and $\mathbf{S_4(\bar{n}, N, \tau)}$

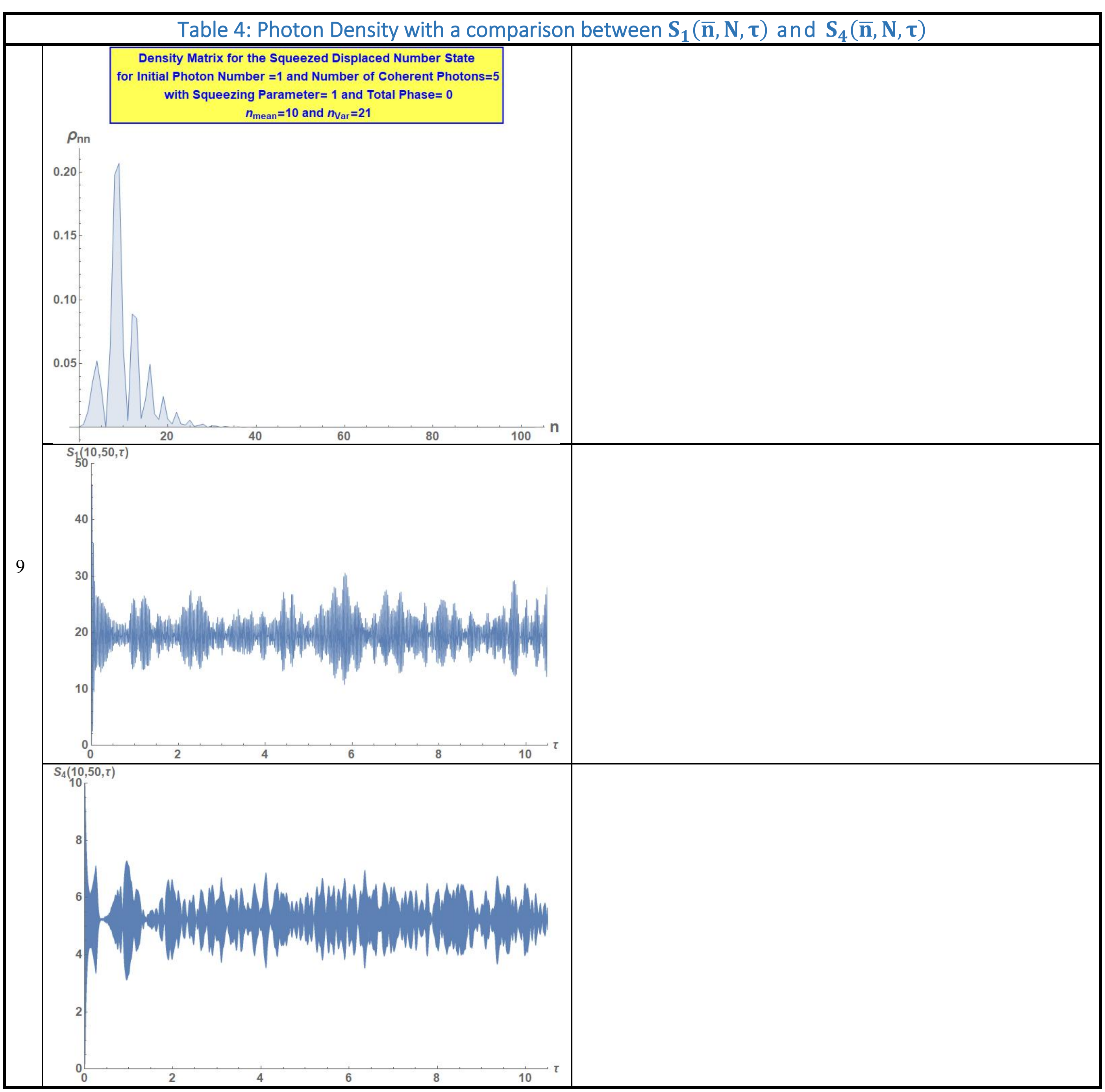

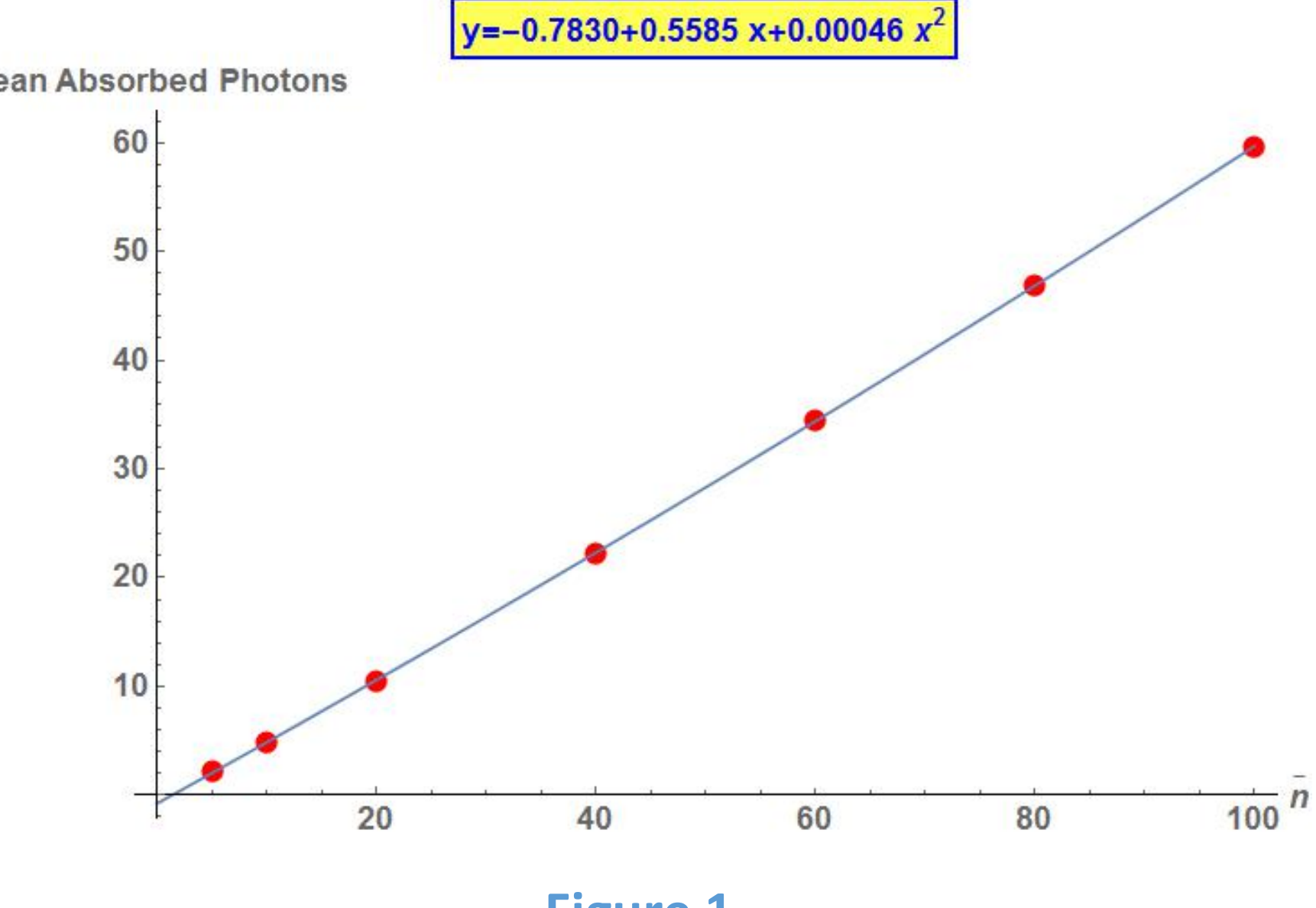

**Figure 1**